\documentclass[prb,aps,amssymb,nofootinbib,twocolumn,showpacs]{revtex4}
\usepackage{amsmath}
\usepackage{amssymb}
\usepackage{amsthm}
\usepackage{amsfonts}
\usepackage{enumerate}
\usepackage{latexsym}
\usepackage[dvips]{graphicx}
\usepackage{float}
\usepackage{subfigure}

\newcommand{\beq}{\begin{equation}}
\newcommand{\eneq}{\end{equation}}
\newcommand{\beqs}{\begin{equation*}}
\newcommand{\eneqs}{\end{equation*}}

\input{epsf}

\begin{document}

\tolerance 1000

\title{ Spinon Deconfinement in Quantum Critical $2+1$ D
Antiferromagnets }

\author { Zaira Nazario$^\star$ and David I. Santiago$^\dagger$ }

\affiliation{$\star$ Max Planck Institute for the Physics of Complex
systems, N\"othnitzer Str. 38, 01187 Dresden, Germany \\
$\dagger$ Instituut-Lorentz for Theoretical Physics, Universiteit Leiden,
P. O. Box 9506, NL-2300 RA Leiden, The Netherlands}

\begin{abstract}

\begin{center}

\parbox{14cm}{ The N\`eel magnetization of $2+1$ D antiferromagnets is
composed of quark-like spin 1/2 constituents, the spinons, as follows
from the $CP^1$ mapping. These quark spinons are confined in both the
N\`eel ordered phase and quantum paramagnetic phases. The confinement
in the quantum paramagnetic phase is understood as arising from
quantum tunneling events, instantons or hedgehog monopole events. In
the present article, we study the approach to the quantum critical
point, where the quantum paramagnetic phase ceases to exist. We find
that irrespective of the intrinsic spin of the antiferromagnet,
instanton events disappear at the deconfined critical point because
instanton tunelling becomes infinitely costly and have zero
probability at the quantum critical point. Berry phase terms relevant
to the paramagnetic phase vanish at the quantum critical point, but
make the confinement length scale diverge more strongly for
half-integer spins, next strongest for odd integer spins, and weakest
for even integer spins. There is an emergent photon at the deconfined
critical point, but the ``semimetallic'' nature of critical spinons
screens such photon making it irrelevant to long distance physics and
the deconfined spinons are strictly free particles. A unique
prediction of having critical free spinons is an anomalous exponent
$\eta$ for the susceptibility exactly equal to one. Experimentally
measurable response functions are calculated from the deconfined
spinon criticality.}

\end{center}
\end{abstract}
\pacs{75.10.-b,75.40.Cx,,75.40.Gb,75.40.-s}
\date{\today}

\maketitle

\section{Introduction}

Shortly after the dawning days of renormalization group
studies\cite{wilson1} of thermodynamic critical phenomena (continuous
finite temperature phase transitions), this work was generalized to
quantum critical phenomena (continuous zero temperature phase
transitions\cite{hertz}) induced by tuning parameters of the
underlying Hamiltonian rather than the temperature. Since the mid
1970s\cite{hertz}, quantum phase transitions have attracted ever
increasing theoretical and experimental activity. Quantum critical
behavior has been obtained from quantum fluctuations of the order
parameter\cite{hertz,millis}. It is then concluded that systems with
$d$ spatial dimensions have quantum critical points identical to
thermal critical points in $d+z$ dimensions when the time direction
scales as $z$ space dimensions. In this traditional approach, the
quantum transition is studied via the Wilson renormalization group in
which fluctuations of the order parameter are taken properly into
account. This is the Landau-Ginzburg-Wilson (LGW) approach. On the
other hand, some measurements can be interpreted as casting doubt on
such a picture\cite{exp2}. In particular, critical exponents are
coming out different than what is predicted. {\it The exponents are
not those of the classical $d + z$ theory with order parameter
fluctuations only.}

There have been recent suggestions\cite{bob1,sachdev2} that there will
be quantum critical physics which do not follow from LGW order
parameter fluctuations alone\cite{bob1,sachdev2}. The new physics
consists of the existence low energy elementary excitations intrinsic
to, and existing only at the critical point, which will contribute and
can modify the quantum critical properties. It was postulated that
these excitations will be fractionalized\cite{bob1,sachdev2}. That
quantum critical points will have unique eigenstates is generally true
as long as the critical propagator has an anomalous exponent. Such
excitations are expected to be fractionalized, but they need not be so
in all cases. These critical degrees of freedom provide critical
fluctuations beyond those of the order parameter fluctuations which
are usually included in the standard Ginzburg-Landau-Wilson (LGW)
phase transition lore.

There are aspects of continuous phase transitions universal to both
classical thermodynamic criticality and quantum mechanical
criticality. Both types of transitions are characterized by a
diverging length scale as it is impossible for a macroscopic system to
qualitatively change behavior unless there are arbitrarily large scale
fluctuations or correlations, either thermal, quantum or
both\cite{wilson1, hertz, sachdev1}. This diverging length scale makes
the critical properties universal and independent of microscopic
details, except for the most general details like symmetry and
dimensionality. The diverging correlations make the system respond to
external stimuli in a scale invariant manner.

The scale invariance universal to both thermal and quantum transitions
is characterized by critical exponents. To be somewhat more explicit,
we concentrate in relativistic quantum critical points, but we
emphasize that this physics can take place in other systems. For such
a system, which we take to be an antiferromagnet, we are interested in
the N\'eel magnetization Green's function, or staggered magnetic
susceptibility. We will think of a transition between a N\'eel ordered
(antiferromagnetic) and disordered (paramagnetic) phase.  

    \begin{figure}[h!]
      \center
      \includegraphics[width=7cm,height=5cm]{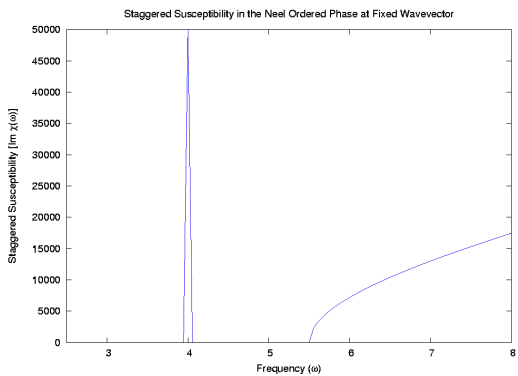}
      \caption{Density of states in the N\'eel ordered phase.}
      \label{propord}
    \end{figure}

In the ordered phase the transverse Green's function or susceptibility
corresponds to spin wave propagation and it has a nonanalyticity in
the form of a pole corresponding to such propagation:
    \beq
      \langle \vec n(-\omega, -\vec k) \cdot \vec n(\omega, \vec
      k)\rangle = \frac{Z(\omega, \vec k)}{c^2k^2 - \omega^2} +
      G_{\text{incoh}} (\omega, \vec k) \,.
    \eneq
Here $Z(\omega, \vec k)$ is between 0 and 1, and the incoherent
background $G_{\text{incoh}}$ vanishes at long wavelengths and small
frequencies. The pole structure of the Green's function is clearly
illustrated when one plots the imaginary part of the N\'eel
magnetization propagator as shown in figure \ref{propord}. The fact
that the Green's function has a pole means that transverse Goldstone
spin waves are low energy eigenstates of the antiferromagnet. At
criticality, the system has no N\'eel order and thus Goldstones cannot
be elementary excitations of the system.
    \begin{figure}[!h]
      \center
      \includegraphics[width=7cm,height=5cm]{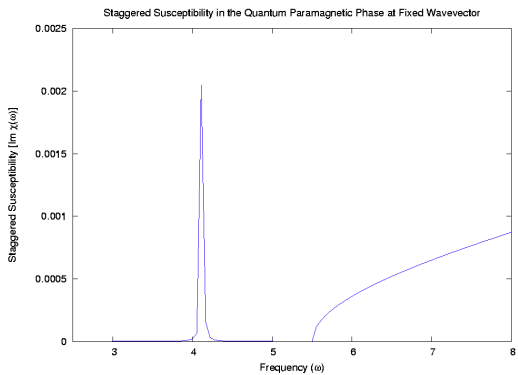}
      \caption{Density of states in the quantum paramagnetic phase.}
      \label{propdis}
    \end{figure}

In the disordered phase the Green function or susceptibility
corresponds to spin wave propagation with all three polarizations and
it has a pole nonanalyticity corresponding to such propagation:
    \beq
      \langle \vec n(-\omega, -\vec k) \cdot \vec n(\omega, \vec
      k)\rangle = \frac{A(\omega, \vec k)}{c^2k^2 + \Delta^2-
      \omega^2} + G_{\text{incoh}} (\omega, \vec k) \,.
    \eneq
Here $A(\omega, \vec k)$ is between 0 and 1, and the incoherent
background $G_{\text{incoh}}$ vanishes at long wavelengths and small
frequencies, $\Delta$ is the gap to excitations in the disordered
phase. The pole structure of the Green's function is clearly
illustrated when one plots the imaginary part of the N\'eel
magnetization propagator as shown in figure \ref{propdis}. That this
Green's function has a pole means that triplet or triplon spin waves
are low energy eigenstates of the disordered antiferromagnet. For 2+1
D antiferromagnets, and in general for antiferromagnets below the
upper critical dimension, the quasiparticle pole residue $A$ vanishes
as the system is tuned to the quantum critical point\cite{halp,
sachdev3}. At criticality, triplon excitations have no spectral weight
and thus triplons cannot be elementary excitations of the system.
    \begin{figure}[!h]
      \center 
      \includegraphics[width=7cm,height=5cm]{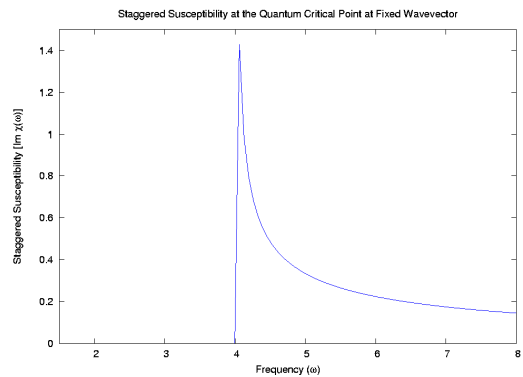}
      \caption{Density of states at the quantum critical point.}
      \label{propcrit}
    \end{figure}

On the other hand right at criticality the response function below the
upper critical dimension (below which $\eta \neq 0$, while above
$\eta=0$) has nonanalyticities that are worse than poles
    \beq
      \langle \vec n(-\omega, -\vec k) \cdot \vec n(\omega, \vec
      k)\rangle = A'\left(\frac{1}{c^2k^2 - \omega^2} \right)^{1 -
      \eta/2}
    \eneq
as obtained from the renormalization group studies of the nonlinear
sigma model\cite{polya2, br1, br2, halp}. Below the upper critical
dimension $\eta$ is a nonintegral universal number for each
dimensionality. This critical susceptibility has no pole structure,
but has a branch cut. It sharply diverges at $\omega=ck$ and has an
imaginary part for $\omega>ck$. The nonanalytic structure of the
Green's function is clearly illustrated when one plots the imaginary
part of the N\'eel magnetization propagator as shown in figure
\ref{propcrit}.  Branch cuts in quantum many-body or field theory
represent immediate decay of the quantity whose Green function is
being evaluated. Hence the elementary excitations or eigenstates of
the noncritical quantum mechanical phases break up as soon as they are
produced when the system is tuned to criticality: they do not have
integrity. The complete lack of pole structure and the branch cut
singularity below the upper critical dimension mean that the
elementary excitations of the quantum mechanical phases away from
criticality, the spin waves, {\it cannot even be approximate
eigenstates} at criticality as they are absolutely unstable.

The quantum critical point is a unique quantum mechanical phase of
matter, which under any small perturbation becomes one of the phases
it separates. It is a repulsive fixed point of the renormalization
group. As far as the transition from one quantum mechanical phase to
the other is continuous, and both phases have different physical
properties, the critical point will have its unique physical
properties different from the phases it separates. The properties of
the critical point follow from the critical Hamiltonian $H(g_c)$
($g_c$ is the critical coupling constant), which will have a unique
ground state and a collection of low energy eigenstates which are its
elementary excitations. These low energy eigenstates are different
from those of each of the phases as long as we are below the upper
critical dimension. {\it As a matter of principle, all quantum
critical points below the upper critical dimension will have their
intrinsic elementary excitations}.

We have seen that below the upper critical dimension, the excitations
of the stable quantum phases of the system become absolutely unstable
and decay when the system is tuned to criticality. The question comes
to mind immediately: what could they be decaying into? When one tries
to create an elementary excitation of one of the phases, it will decay
immediately into the elementary excitations of the critical point. The
critical excitations will be bound states of the excitations of the
stable phases the critical point separates. These bound states could
be fractionalized as conjectured by Laughlin\cite{bob1} and Senthil,
{\it et. al.}\cite{sachdev2}, but they need not be in all cases. These
critical degrees of freedom are responsible for corrections to the LGW
phase transition canon\cite{sachdev2}. {\it The intrinsic quantum
critical excitations contribute to the thermodynamical and/or physical
properties of the quantum critical system.}

In the present work, we develop these ideas to find if they occur in
antiferromagnets. We will consider the disordered paramagnetic phase
of $2+1$ D antiferromagnets and the approach to the quantum critical
point from such a phase. These are described by the nonlinear sigma
model augmented by Berry phase terms as originally discovered by
Haldane\cite{hal1,hal2} and developed by
others\cite{vector,nohopf,subir}:
    \beq
      \mathcal Z =\int \mathcal D\vec n \delta(\vec n^2-1)
      e^{-S/\hbar}
    \eneq
with
    \beq
      S = S_{B} + \int_0^\beta \! \frac{d(c\tau)}{2ga} \! \int \!
      d^2\vec r \left[(\nabla_{\vec r}\vec n)^2 +
      \frac{1}{c^2}(\partial_\tau\vec n)^2 \right] \,.
    \eneq
$g = 2\sqrt{2} / S$ is the dimensionless coupling constant, $a$ is the
lattice constant, $S$ is the microscopic spin with $\hbar$ included
(not to be confused with the Euclidean action), and $S_B$ is the Berry
phase term. Their effect is nonzero only in the disordered phase of
the Heisenberg antiferromagnet as first suggested by
Haldane\cite{hal2} and worked out by Read and
Sachdev\cite{sachread}. In particular, when the microscopic spins in
the lattice are half-odd integers, the paramagnetic ground state has a
spin Peierls bond order that breaks the lattice symmetry and is
four-fold degenerate. For odd integer spins the paramagnetic ground
state has a spin Peierls bond order that breaks the lattice symmetry
and is two-fold degenerate. For even integer spins we have a valence
bond solid that does not break the lattice symmetry. All of these
quantum paramagnetic phases have a spin triplet gapped excitation, the
triplon, and a spin zero gapped collective mode.

Motivated by apparent anomalies and interesting effects in the physics
of cuprate superconductors, whose parent state is a Mott insulating $2
+ 1$ D antiferromagnet, Laughlin suggested\cite{bob1} that the N\'eel
field would break up into constituents ``quark'' spinons at the
quantum critical point between a N\'eel ordered and a quantum
paramagnetic phase. A couple of years ago, Fisher, Sachdev and
collaborators\cite{sachdev2} suggested that such a physics indeed
occurs, but only in spin $1/2$, $2+1$ D antiferromagnets.

The quantum paramagnetic phase of antiferromagnets is equivalent to a
``charged'' $CP^1$ spinon field whose fictitious $U(1)$ charge couples
to an emergent $U(1)$ gauge field or photon generated by the
fluctuations of the $CP^1$ field\cite{divecchia, witten}. The $CP^1$
mapping ($\vec n = z^\dagger \vec \sigma z$, with $\vec \sigma$ the
vector of sigma matrices) is obtained from the nonlinear sigma model
description of antiferromagnets in terms of the N\'eel field $\vec
n$\cite{hal1}. Haldane discovered that in $1+1$ and $2+1$ D, the
nonlinear sigma model needs to be augmented by Berry phase terms in
the paramagnetic or spin disordered phase\cite{hal1,hal2,sachread},
which in $2+1$ D leads to breaking of lattice symmetries for odd
integers and half odd integer spins\cite{hal2,sachread}. If one doubts
the nonlinear sigma model mapping of antiferromagnets in the
disordered phase, Read and Sachdev\cite{sachread}, starting from the
Heisenberg model, showed that the disordered phase is indeed described
by ``charged'' $CP^1$ ``quark'' spinons $z$ coupled to an emergent
photon $A_\mu$.

In $2+1$ D, the $CP^1$ model has important tunneling
events\cite{murthsach} which correspond to instanton hedgehog events
that effectively make the $U(1)$ gauge field compact. Moreover, if we
start from the appropriate Heisenberg lattice description, the gauge
group is necessarily compact. Polyakov showed\cite{polyaqed,polyab}
that compact QED confines as the Wilson loop\cite{wilson} obeys an
area law when instanton or monopole events are included. Polyakov's
proof did not include matter, but it is believed\cite{sachread}, and
we show below, that the presence of charged matter does not eliminate
the tunneling instanton events {\it as long as the charged matter is
massive}. Therefore, spinons are confined in the paramagnetic phase
and are thus closely analogous to the quarks of Quantum
Chromodynamics.

Fisher, Sachdev and collaborators suggested that the Berry
phase-induced quadrupling of instanton events intrinsic to spin $1/2$
antiferromagnets makes such monopole events irrelevant at the critical
point between the paramagnetic and N\'eel ordered
phases\cite{sachdev2}. Therefore, spinons are deconfined at such a
critical point for spin $1/2$ antiferromagnets. They further suggested
that at the deconfined quantum critical point the critical exponent
$\eta$ of the N\'eel field correlator is due to the decay of the
N\'eel field into the deconfined quark spinons. These spinons will be
the intrinsic excitations of the quantum critical point.

We study here the question of deconfinement, its consequences and how
it occurs. We find that when deconfinement of the N\'eel field into
two spin $1/2$ quarks occurs, {\it the critical exponent $\eta$ is
exactly equal to 1 regardless of the origin of
deconfinement}. Deconfinement will occur whenever instanton events
vanish. We find that for all values of the microscopic spin, integers
and half-odd integers, instanton events vanish at the quantum critical
point where the paramagnetic phase ceases to exist. The vanishing of
instantons happens for two reasons. The masslessness of the spinons at
criticality screens the instanton fields, making them irrelevant at
long distances. Furthermore, {\it for all values of the microscopic
spins}, the Euclidean action of instanton events becomes infinite at
the quantum critical point due to the masslessness of the
spinons\cite{murthsach}, and thus {\it the probability of instanton
events at criticality is zero. Hence deconfinement occurs independent
of the value of the microscopic spin}. 

On the other hand, there is some dependence on the microscopic spin on
the quantum paramagnetic phase as the lattice symmetry breaking
depends on the spin value through Berry phase terms\cite{sachread}. We
also found further dependence on the microscopic spin as the system is
tuned to the quantum critical point since our analysis also has the
consequence that the confinement length will diverge faster upon
approach to criticality for half-odd integer spins, next fastest for
odd integer spins, and slowest for even integer spins as a consequence
of the Berry phase terms relevant to the quantum paramagnetic phase.

For the first time, we write down the effective critical theory and
from it, calculate experimental consequences. At criticality, there is
still a $U(1)$-mediated gauge interaction between massless spinons, so
in principle they might not be free. We find that at criticality,
spinons are very mobile because of their masslessness. Hence they
screen very effectively their gauge interaction so they are strictly
free at long distances. The decay of the N\'eel field into two free
particles leads to a critical exponent of $\eta=1$ exactly. This
result can be used as a diagnostic of deconfined criticality, that is,
{\it for a free deconfined spinon critical point we predict a critical
exponent $\eta$ exactly equal to one}.

The deconfined critical points studied and elucidated here seem to be different
than the $2 + 1$ D Heisenberg critical points. It has been suggested before
that these two different types of critical points might occur in $2+1$
D\cite{sachdev2}. One particular suggestion is that interactions irrelevant
to the N\'eel and quantum paramagnetic phases turn the Heisenberg
critical point into a deconfined critical point and there seems to be indirect
numerical evidence for such physics\cite{sandvik}. Before making strong
conclusions one must wait for experimental evidence and/or further and more
explicit numerical evidence.

\section{$CP^1$ mapping of the $O(3)$ nonlinear sigma model}

We consider the $CP^1$ mapping of $SO(3)$ vectors $\vec
n$\cite{witten,divecchia}
    \beq
      \vec n = z^\dagger \vec\sigma z
    \eneq
where
    \beq
      z = \left(
      \begin{matrix}
	z_1 \\
	z_2
      \end{matrix} \right)
    \eneq
with the restriction 
    \beq
      |z|^2 \equiv |z_1|^2 + |z_2|^2 = 1
    \eneq
inherited from $\vec n^2 = 1$. The $z$'s are bosonic. Notice that the
$CP^1$ map has 4 variables. The restriction $|z|^2 = 1$ eliminates
one, leaving 3 independent variables. This seems to be one too much as
the $O(3)$ nonlinear sigma model has only 2 independent variables
since one is eliminated by the nonlinear condition. This is not so as
one of the variables is redundant because of the gauge symmetry $z
\rightarrow z e^{i\theta}$. The $z$'s are spinors, i.e. spin $1/2$
objects. The $z$'s are the quark-like spinon constituents of the
N\'eel field $\vec n$.

With this mapping the $O(3)$ nonlinear sigma model partition function
becomes
    \beq
      \mathcal Z = \int \mathcal D z \mathcal D z^\dagger \delta
      (|z|^2 - 1) e^{-S}
    \eneq
with Euclidean action
    \beq
      S = \frac{2}{ga} \int d^3r \left( |\partial_\mu z|^2 -
      |z^\dagger \partial_\mu z|^2 \right) + S_B \,.
    \eneq
and Berry phase
    \beq
      S_B = \sum_i \epsilon_i \int_0^\beta d\tau z^\dagger
      \partial_\tau z \;. 
    \eneq
The lattice regularization is necessary to define the Berry phase, as
the Berry phase has microscopic lattice sensitivity. We can decouple
the quartic term via a Hubbard-Stratonovich transformation leading
to\cite{divecchia,witten}
    \beq
      \mathcal Z = \int \mathcal D z \mathcal D z^\dagger \mathcal D
      A_\mu \delta (|z|^2 - 1) e^{-S}
    \eneq
    \beq
      S = \frac{2}{ga} \int d^3 r |(\partial_\mu - i A_\mu) z|^2 + S_B
    \eneq
Now the gauge invariance is explicit as the kinetic term for the
$CP^1$ fields is built up of covariant derivatives. Since the action
is quadratic in $A_\mu$, the saddle point evaluation about the minimum
with respect to $A_\mu$ is exact and can be used to go back to the
original action by substituting the solution to the $A_\mu$ equation
of motion
    \beq
      A_\mu = \frac{i}{2} \left[z \partial_\mu z^\dagger - z^\dagger
      \partial_\mu z\right] = i z\partial_\mu z^\dagger = -i z^\dagger
      \partial_\mu z
    \eneq

In $2+1$ dimensions, which is our main concern here, the Berry phase
terms can be written in terms of the gauge fields as\cite{sachread}
    \beq
      S_B = \sum_s i \pi S \zeta_s q_s
    \eneq
where $S$ is the microscopic spin (not to be confused with the
Euclidean action), $s$ are points in the dual lattice around which a
hedgehog or magnetic monopole of strength $q_s$ is centered and
$\zeta_s$ is 0, 1, 2 and 3 depending on which dual lattice, $W,X,Y,Z$,
the monopole is centered.
    \begin{center}
      \includegraphics[width=6cm,height=6cm]{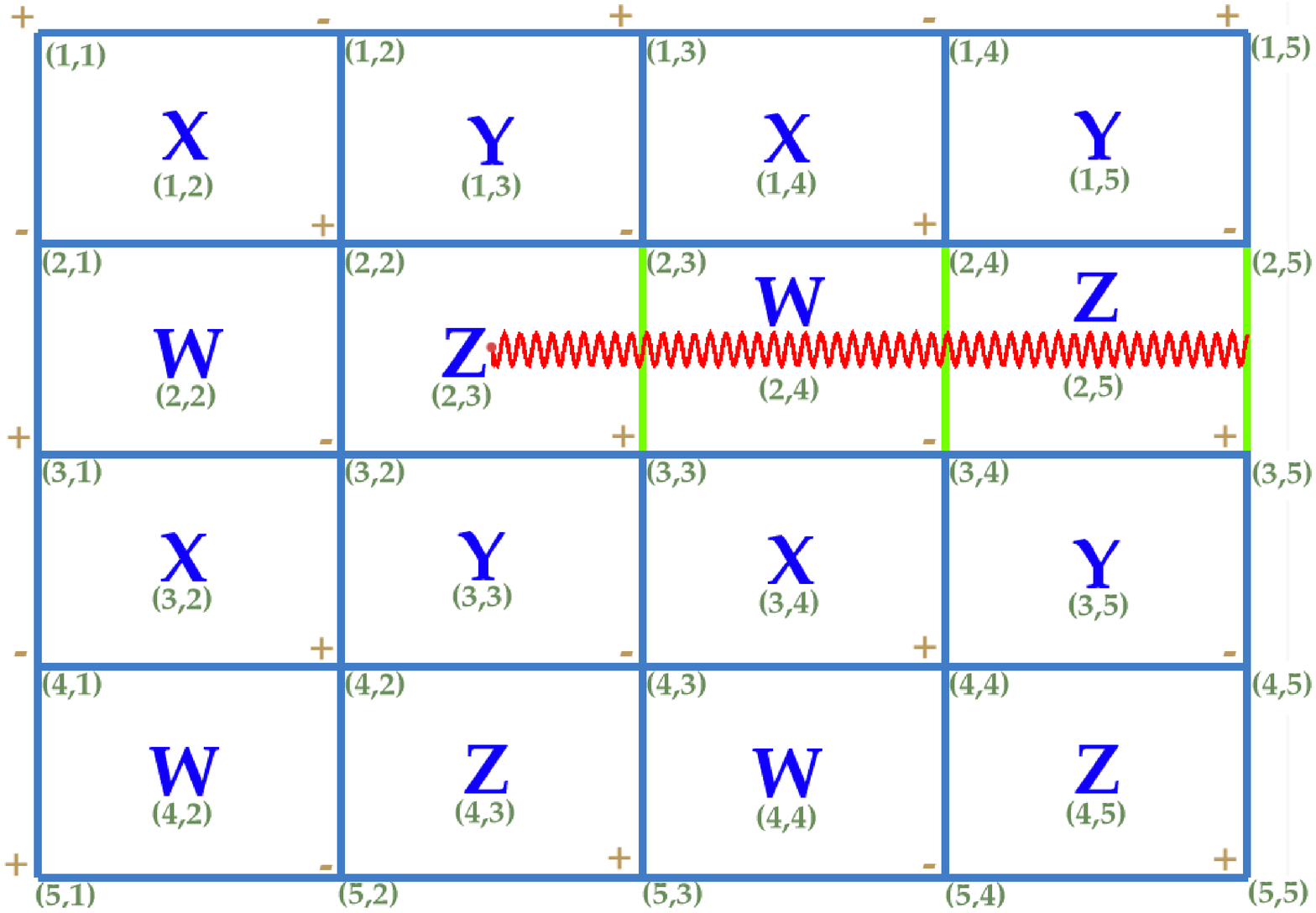}
    \end{center}
$q_s$ is given by
    \beq
      q_s = \frac{1}{2\pi}\int_\Sigma dS_{\mu\nu} F_{\mu\nu}
    \eneq
where the normal to the surface $\Sigma$ is closed and $F_{\mu\nu}$ is
the Maxwell tensor for $A_\mu$. Such spinon induced gauge field
tunneling event cannot happen unless the gauge field is compact. Some
readers might object that the nonlinear sigma model from which we
started should not apply in the paramagnetic phase. For those readers
we point out that one can start from the Heisenberg model and use a
Schwinger boson representation on the lattice and one obtains a
similar action, and leads to the same $CP^1$ model with the advantage
that in such a derivation the emergent gauge field is necessarily
compact. In fact, such a derivation was performed by Read and
Sachdev\cite{sachread}. 

We can write the delta function in the partition function in its
integral form, so that
    \beq
      \mathcal Z = \int \mathcal D z^\dagger \mathcal D z \mathcal D
      A_\mu \mathcal D \lambda e^{-S}
    \eneq
where now
    \begin{align}
      \begin{aligned}
	S &= \frac{2}{ga} \int d^3r \, \left\{ |(\partial_\mu - i
	A_\mu) z|^2 + i \lambda \frac{g a}{2} (z^\dagger z - 1)
	\right\} \\
	&+ S_B \;.
      \end{aligned}
    \end{align}
Integrating the $z$'s we obtain an effective action for the $A$'s and
the $\lambda$'s. This effective action has a minimum at $A = 0$ and
$\lambda =$ constant. Varying the effective action with $A = 0$ and
$\lambda =$ constant with respect to $\lambda$ gives a self
consistency equation for $\lambda$
    \beq
      1 - \frac{4\pi^2}{g} = \frac{m}{\Lambda}
      \arctan\left(\frac{\Lambda^2}{m^2}\right)
    \eneq
where $\Lambda = 1/a$ is the inverse lattice constant and $\langle
i\lambda\rangle = 2m^2 / (ga)$. Since $\lambda$ is a mass term for the
$z$'s, the nonlinear condition $|z|^2 = 1$ generates a mass $m$ for
the spinons $z$ fields for $g \ge g_c \equiv 4\pi^2$. $g_c$
corresponds to the quantum critical point where the quantum
paramagnetic phase dies and in all probability gives rise to the
N\'eel ordered phase.

The $A_\mu$ seems to play a passive role, as it has no dynamics of its
own in the bare action. This is not so once fluctuations and dynamical
corrections are calculated. The gauge field acquires dynamics through
the $z$ field fluctuations leading to an effective action with a
Maxwell action for the gauge fields. In fact the long wavelength gauge
field propagator gives\cite{witten}
    \beq
      \langle A_\mu (-k_\alpha) A_\nu (k_\alpha)
      \rangle_{1-\text{loop}} = \frac{2}{3\pi m} (k_\mu k_\nu -
      \delta_{\mu\nu} k^2) \,.
    \eneq
This term is the same as would be obtained from a term in the action
of the form 
    \beq
      \frac{1}{3\pi m} \int d^3 r F_{\mu\nu}^2 \equiv \frac{1}{4e^2}
      \int d^3 r F_{\mu\nu}^2 \,.
    \eneq
Hence $e^2 \simeq 3m\pi/4$ in two dimensions. Higher order corrections
will renormalize $e^2$ such that $e^2 = m \, h(g,m/\Lambda)$, where
$h$ is dimensionless. Higher order corrections also generate terms
that vanish at long wavelengths and are thus dropped. Therefore, the
effective action is one for massive spinon fields $z$ coupled to a
$U(1)$ gauge field
    \begin{align}
      \begin{aligned}
	S &= \frac{2}{ga} \int d^3 r \left\{ |(\partial_\mu - i
	A_\mu)z|^2 + m^2 |z|^2 \right. \\
	&+ \left. i \delta \lambda \frac{ga}{2} (|z|^2 - 1)\right\} +
	\int \frac{d^3 r}{4e^2} F_{\mu\nu}^2 + S_B \;.
      \end{aligned}
    \end{align}
We then have a theory of ``charged'' spinons coupled to an emergent
compact $U(1)$ gauge field or photon\cite{divecchia,witten,sachread}.

\section{Effective Action for the Quantum Paramagnetic Phase of 
$2+1$ D Antiferromagnets}

For the paramagnetic or disordered phase of $2+1$ D antiferromagnets,
$g>g_c$, we have seen that the spinon fields $z$ acquire a gap. The
quantum critical point corresponds to the spinons becoming massless at
$g = g_c$. We also saw that the adiabatic fluctuations of the spinons
generate dynamics for the gauge fields. We have an emergent compact
$U(1)$ gauge field coupled to a complex spinor representing the spinon
fields. In our case the gauge field is compact as follows from the
lattice Schwinger boson derivation of the effective $CP^1$ theory for
the Heisenberg model. Therefore, the quantum paramagnetic phase and
quantum critical point of $2+1$ D antiferromagnets is mapped to
compact $2+1$ D QED with two complex matter fields with an $SU(2)$
internal symmetry. The internal symmetry is effectively the original
invariance of the nonlinear sigma model and plays a passive role for
the most interesting properties of the paramagnetic phase to be
developed in this section.

If the emergent photon did not correspond to a compact $U(1)$ group,
the paramagnetic phase would have massive spinon excitations and
massless photons. On the other hand, on careful thought, this seems
completely wrong as the paramagnetic phase is fully gapped and the
emergent photon would be gapless. It turns out that compact QED in
$2+1$ D, as originally shown by Polyakov\cite{polyaqed,polyab}, has
monopole tunneling events, usually called instantons or hedgehogs,
such that the ground state of the theory is a monopole
condensate. Such a ground state gives the photon a mass {\it without
breaking the gauge symmetry} and makes the Wilson loop obey an area
rather than a perimeter law\cite{wilson}, rendering the theory
confining. Polyakov's proof did not include matter. We show below that
including the charged matter spinon fields, the theory still confines
and the spectrum is fully massive and singlet with respect to the
$U(1)$ gauge field.

Before moving to study the effects of instantons, we remind the reader
that we do not have pure QED, but we have the gauge fields coupled to
charge spinon fields $z$, which satisfy the nonlinear condition $|z|^2
= 1$ as enforced by the Lagrange multiplier $\delta\lambda$. In order
to find the effect of the spinons on the theory, they can be
integrated out of the partition function explicitly to yield
    \begin{align}
      \begin{aligned}
	Z &= \int \mathcal D (\delta\lambda) \mathcal D A_\mu e^{-S}
	\\
	&\times \text{det}^{-2} \left[ - \frac{2}{ga} (\partial_\mu -
	i A_\mu)^2 + i\delta\lambda + \frac{2}{ga} m^2 \right]
      \end{aligned}
    \end{align}
with
    \beq
	S = \int d^3 r \left[ \frac{1}{4e^2} F_{\mu\nu}^2 -
	i\delta\lambda \right] + S_B \,.
    \eneq
Using $\text{det} M = \exp{\left[\text{tr} \ln M \right]}$, the
partition function can be rewritten as
    \beq
      Z = \int \mathcal D (\delta\lambda) \mathcal D A_\mu
      e^{-S_{eff}}
    \eneq
with
    \begin{align}
      \begin{aligned}
	S_{eff} &= S_B + \int d^3 r \left[ \frac{1}{4e^2} F_{\mu\nu}^2
	- i\delta\lambda \right] \\
	&+ 2 \text{tr} \ln \left[ - \frac{2}{ga} (\partial_\mu - i
	A_\mu)^2 + i\delta\lambda + \frac{2}{ga} m^2 \right] \;.
      \end{aligned}
    \end{align}

For the moment we only consider the $\delta\lambda$ dependent part of
the action and its path integration. We can expand about the saddle
point $\delta\lambda = 0$ and path integrate over $\delta\lambda$ to
obtain
    \beq
      \text{det} M \int \mathcal D (\delta\lambda) \, \exp{\left[ -
      \delta\lambda M \delta\lambda \right]} = \text{det} M^{1/2}
    \eneq
where the operator $M$ is the inverse of the square of the operator $-
2 (\partial_\mu - iA_\mu)^2 / (ga) + 2 m^2 / (ga)$. Hence the total
partition function after taking care of the constraint, i.e. after
integrating over $\delta\lambda$, is
    \begin{align}
      \begin{aligned}
	Z &= \int \mathcal D A_\mu e^{-S_{f1}} \\
	S_{f1} &= S_B + \int \frac{d^3r}{4e^2} F_{\mu\nu}^2 \\
	&+ \text{tr} \ln \left[ - \frac{2}{ga} (\partial_\mu -
	iA_\mu)^2 + \frac{2}{ga} m^2 \right] \;.
      \end{aligned}
    \end{align}
This is equivalent to 
    \begin{align}
      \nonumber
      Z &= \int \mathcal D A_\mu \text{det}^{-1} \left[ - \frac{2}{ga}
      (\partial_\mu - iA_\mu)^2 + \frac{2}{ga} m^2 \right] e^{-S_{f2}}
      \\
      S_{f2} &= \int \frac{d^3r}{4e^2} F_{\mu\nu}^2 + S_B
    \end{align}
or disentangling the determinant in terms of a complex scalar field
$\Phi$
    \begin{align}
      \begin{aligned}
	Z &= \int \mathcal D A_\mu e^{-S_{f3}} \\
	S_{f3} &= \frac{2}{ga} \int d^3r \left[ \Big | (\partial_\mu -
	iA_\mu) \Phi \Big |^2 + m^2 \Phi^* \Phi \right] \\
	&+ \frac{1}{4e^2} \int d^3r F_{\mu\nu}^2 + S_B \,.
      \end{aligned}
    \end{align}
Hence antiferromagnets in $2+1$ D, which were shown to be equivalent
to compact QED coupled to spinor spinon fields $z$ with the constraint
$|z|^2 = 1$ become equivalent to $2+1$ D compact QED coupled to
charged complex scalar fields once the constraint is enforced. 

Now the complex scalar fields can be integrated out to yield their
effect on the theory. When they are integrated out their main effect
is to screen electromagnetic fields by a factor $\epsilon_m$ which at
long distances takes the form
    \beq
      \epsilon_m = 1 + \frac{e^2}{m} \, f\left( \frac{m}{\Lambda}\,,
      \frac{e^2}{\Lambda} \right)
    \eneq
where $f > 0$. The largest contributions to the screening effects of
the $\Phi$, and hence the spinons, at long distances can be easily
calculated by computing and summing polarization diagrams. To one loop
order, in the Landau gauge, we obtain for the long wavelength photon
propagator
    \beq
      \langle A_\mu (-k) A_\nu (k) \rangle \simeq \frac{1}{k^2}
      \left( 1 - \frac{e^2}{3\pi m} \right) \left( \delta_{\mu\nu} -
      \frac{k_\mu k_\nu}{k^2} \right)
    \eneq
Hence we see that the spinons decrease the effective magnetic charges
and thus have a diamagnetic screening effect which we parametrize by a
screening constant $\epsilon_m$. The most important diagrams
contributing to this screening are the bubble diagrams, which we sum
to obtain
    \begin{align}
      \begin{aligned}
	&\langle A_\mu (-k) A_\nu (k) \rangle = \frac{1}{\epsilon_m
	k^2} \left( \delta_{\mu\nu} - \frac{k_\mu k_\nu}{k^2} \right)
	\\
	&\simeq \frac{1}{\left[ 1 + e^2/(3\pi m) \right] k^2} \left(
	\delta_{\mu\nu} - \frac{k_\mu k_\nu}{k^2} \right) \;.
      \end{aligned}
    \end{align}
Therefore in the random phase approximation (RPA), $\epsilon_m = 1 +
e^2/(3\pi m)$. We finally obtain that the effective action including
the effects of spinons is given by screening electromagnetic fields by
the appropriate factor $\epsilon_m$
    \beq
      S = \frac{1}{4 e^2 \epsilon_m} \int d^3r F_{\mu\nu}^2 + S_B \,.
    \eneq

\section{Confinement of Spinons in the Quantum Paramagnetic Phase 
of $2+1$ D Antiferromagnets}

Now that we have seen that once the effects of the spinons is
included, the effective action for the antiferromagnet is equivalent
to compact QED with an appropriate screening factor induced by the
spinons, we need to take into account the nontrivial tunneling effects
arising from the compactness of the gauge
group\cite{polyaqed,polyab,sachread}. The monopole tunneling events
considered by Polyakov create electromagnetic fields according to
    \beq
      \nabla \times \vec H = 0 \;, \qquad \nabla \cdot \vec H = 2\pi q
      \delta(\vec r - \vec r')
    \eneq
where
    \beq
      H_\mu = \frac{1}{2} \epsilon_{\mu\nu\lambda} F_{\nu\lambda}
    \eneq
and $q$ is necessarily an integer as follows from the compactness of
the gauge field. These tunneling events correspond to certain
tunneling events for the spinons via the mapping
    \beq
      A_\mu = \frac{i}{2} \left[z \partial_\mu z^\dagger - z^\dagger
      \partial_\mu z\right] = i z\partial_\mu z^\dagger = -i z^\dagger
      \partial_\mu z
    \eneq
In fact, such spinon tunneling events have been calculated by Murthy
and Sachdev\cite{murthsach} and found to have Euclidean action given
by
    \beq
      S_{cl} = 2\rho (q) \ln \left(\Lambda/m \right)
    \eneq
leading to the instanton fugacity
    \beq
      \zeta = e^{-S_{cl}} = \exp{\left[-2\rho (q) \ln \left(\Lambda/m
      \right) \right]} = \left[ \frac{m}{\Lambda}\right]^{2\rho (q)}
      \;.
    \eneq
This instanton action depends on the charge $q$ of the tunneling
event. For single instanton event $\rho \equiv \rho(1) \simeq
0.06229609$. 

From the form above we see that the instanton events would have a
Coulomb law field
    \beq
      H_\mu = \sum_a \frac{q_a \left[ r_\mu - r_\mu^a \right]}{2 |r -
      r^a|^3}
    \eneq
where $q_a$ are the charges of the tunneling events centered at
$r_\mu^a$ in $2+1$ D space-time. If $A_{cl}^\mu$ are the classical
gauge fields corresponding to such instanton configurations, a
semiclassical expansion around such configurations with gauge fields
$A_{cl}^\mu + A^\mu$ yields the effective Maxwell action, Berry phases
and instanton Coulomb gas action
    \begin{align}
      \begin{aligned}
	&\left[ \int d^3r \frac{1}{4 e^2 \epsilon_m} F_{\mu\nu}^2
	\right] + i\sum_s \pi S \zeta_s q_s \\
	&+ \frac{\pi}{2 e^2 \epsilon_m} \sum_{s\neq t} \frac{q_s
	q_t}{\left[ (\vec R_s - \vec R_t)^2 + (\tilde \tau_s - \tilde
	\tau_t)^2 \right]^{1/2}}
      \end{aligned}
    \end{align}
where $F_{\mu\nu}$ is the Maxwell field of $A_\mu$. In order to get
the partition function of the antiferromagnet, we have to integrate
over the Euclidean times of the instanton events and sum over the
instanton events centered in the dual lattice. The partition function
for our system is given by
    \beq
      \mathcal Z = \int \mathcal D z \mathcal D z^\dagger \mathcal D
      A_\mu e^{-S_{zA}} \, \mathcal Z_{inst}
    \eneq
where the Euclidean non-instanton action is
    \beq
      S_{zA} = \frac{1}{4 e^2 \epsilon_m} \int d^3r F_{\mu\nu}^2
    \eneq
and
    \beq
      \mathcal Z_{inst} = \sum_{K, q_s} \frac{\zeta^K}{K!}
      \prod_{s=1}^K \left[ \sum_{R_s} \int_0^{\beta}
      \frac{d\tau_s}{a} \right] e^{-S_{inst}}
    \eneq
with Euclidean instanton action
    \begin{align}
      \begin{aligned}
	S_{inst} &= \frac{\pi}{2 e^2 \epsilon_m} \sum_{s\neq t}
	\frac{q_s q_t}{\left[ (\vec R_s - \vec R_t)^2 + (\tilde \tau_s
	- \tilde \tau_t)^2 \right]^{1/2}} \\
	&+ i\sum_s \pi S \zeta_s q_s \;.
      \end{aligned}
    \end{align}
Note that this is a Coulomb gas action augmented by Berry phases.

\subsection{Mapping of the Instanton Coulomb Gas to Sine-Gordon 
Theory Including the Effects of Berry Phases}

As we have just seen, the instanton contributions will provide
constant average field fluctuations in the ground state, and this is
described by a Coulomb gas partition function. This Coulomb gas can be
disentangled by using the fact that the Laplacian operator is the
inverse of the Coulomb potential. Following closely the methods of
Polyakov, and Read and Sachdev, which included the effects of Berry
phases for the first time\cite{polyaqed,polyab,sachread}, we obtain
    \begin{widetext}
      \begin{align}
	\nonumber
	&\mathcal Z_{inst} = \\
	\nonumber
	&\int \mathcal D \chi \left[ \exp\left\{-\frac{e^2
	\epsilon_m}{8\pi} \int_0^\beta d\tau \left[ \sum_{\langle s,t
	\rangle} \left( \chi_s - \chi_t \right)^2 + \sum_s a^2 \left[
	\frac{\partial \chi_s}{\partial \tau} \right]^2 \right]
	\right\} \sum_K \frac{\zeta^K}{K!} \prod_{s=1}^K
	\int_0^{\beta} \frac{d\tau_s}{a} \sum_{R_s, q_s} \exp\left[
	i\sum_{q_s} \left[\pi S \zeta_s + \chi_s \right] q_s \right]
	\right] \\
	\nonumber
	&= \int \mathcal D \chi \left[ \exp\left\{-\frac{e^2
	\epsilon_m}{8\pi^2} \int_0^\beta d\tau \left[ \sum_{\langle
	s,t \rangle} \left( \chi_s - \chi_t\right)^2 + \sum_s a^2
	\left[ \frac{\partial \chi_s}{\partial \tau} \right]^2 \right]
	\right\} \sum_K \frac{\zeta^K}{K!}  \left( \int_0^\beta
	\frac{d\tau_s}{a} \sum_{R_s} 2 \cos \left[ \pi S \zeta_s +
	\chi_s \right] \right)^K \right] \\  
	&= \int \mathcal D \chi \exp\left\{-\frac{e^2
	\epsilon_m}{8\pi^2} \int_0^\beta d\tau \left[ \sum_{\langle
	s,t \rangle} \left( \chi_s - \chi_t \right)^2 + \sum_s a^2
	\left[ \frac{\partial \chi_s}{\partial \tau} \right]^2 \right]
	+ 2 \zeta \int_0^\beta \frac{d\tau_s}{a} \sum_{R_s} \cos
	\left[ \pi S \zeta_s + \chi_s \right] \right\} \\
	\nonumber
	&= \int \mathcal D \chi \exp\left\{-\frac{e^2 \epsilon_m
	a^2}{8\pi^2} \int_0^\beta d\tau \left[ \sum_{\langle s,t
	\rangle} \left( \frac{\chi_s - \chi_t}{a} \right)^2 + \sum_s
	\left[ \frac{\partial \chi_s}{\partial \tau} \right]^2 -
	\frac{16\pi^2}{e^2 \epsilon_m} \frac{(m a)^{2\rho}}{a^3}
	\sum_s \cos \left[ \pi S \zeta_s + \chi_s \right] \right]
	\right\}
      \end{align} 
    \end{widetext}
where $\rho = 0.06229609$ and we sum over instantons with charges $q =
0,1,-1$ since higher monopole charges are strongly suppressed as they
have very small fugacity or probability\cite{polyaqed, polyab,
murthsach}. Hence, as found by Polyakov, we see that the Coulomb gas
is equivalent to a Sine-Gordon theory. The only difference is that the
Sine-Gordon theory we obtain, which is centered in the dual lattice,
is frustrated by the Berry phase terms since the argument of the
cosine has the phase shift $\pi S \zeta_s$ with the values $0,\pi S,
2\pi S$ and $3\pi S$ depending on whether the instanton is in the
$W,X,Y$ or $Z$ sublattice\cite{hal2,sachread}.

We first study the case of even integer spin. In such a case the phase
shift $\pi S \zeta_s$ in the cosine of the Sine-Gordon theory obtained
from instanton events is a multiple of $2\pi$ and thus equivalent to
zero. 

We now consider the case of odd integer $S$. In that case the phase
shift of the cosine of the Sine-Gordon theory is 0 for $\chi_s$ in the
$W$ dual sublattice, and an even integer for $\chi_s$ in the $Y$ dual
sublattice. Therefore, the $W$ and $Y$ lattice are equivalent and we
shall call it $Y$. Similarly, for $\chi_s$ in the $X$ or $Z$
sublattice the cosine in the Sine-Gordon have phase shifts of odd
multiples of $\pi$ and thus are equivalent. We shall call them
$X$. The Sine-Gordon action then becomes
    \begin{widetext}
      \beq
	S_{sg} = 2 \frac{e^2 \epsilon_m a^2}{8 \pi^2} \int_0^{\beta}
	d\tau \left\{ \sum_{\langle s,t \rangle} \left( \frac{\chi_s^X
	- \chi_t^Y}{a} \right)^2 + \sum_s \left[ \left(\frac{\partial
	\chi_s^X}{\partial \tau} \right)^2 + M^2 \cos \chi_s^X \right]
	+ \sum_t \left[ \left(\frac{\partial \chi_t^Y}{\partial \tau}
	\right)^2 - M^2 \cos \chi_t^Y \right] \right\} \;.
      \eneq
where $M^2 = \left[ 16\pi^2 (m a)^{2\rho} \right] / \left[ e^2
\epsilon_m a^3 \right]$ Defining
      \beq
        \chi^X = \chi_1 + \chi_2 \;, \qquad \chi^Y = \chi_1 - \chi_2
      \eneq
and going to the continuum limit, the Sine-Gordon action can be
written as
      \beq
        S_{sg} = 2 \frac{e^2 \epsilon_m}{8 \pi^2} \int_0^{\beta} \int
        d^2\vec r \left[ \left( \nabla \chi_1 \right)^2 + 4 \Lambda^2
        \chi_2^2 - 2\sqrt{2} \Lambda \chi_1 \left( \nabla \chi_2
        \right)^2 + \left( \frac{\partial \chi_1}{\partial \tau}
        \right)^2 + \left( \frac{\partial \chi_2}{\partial \tau}
        \right)^2 - M^2 \sin\chi_1 \sin\chi_2 \right] \;.
      \eneq
    \end{widetext}
Since $\chi_2$ has a mass of the order of the cutoff $\Lambda = 1/a$,
all gradients and time derivatives of $\chi_2$ are suppressed to
zero. We must still minimize the action with respect to $\chi_2$. The
minimization gives
    \beq
      \chi_2 = \frac{M^2}{8\Lambda^2} \sin\chi_1 \cos\chi_2 \simeq
      \frac{M^2}{8\Lambda^2} \sin\chi_1 + \mathcal O\left(
      \frac{M^6}{\Lambda^6} \right)
    \eneq
which, when substituted into the Sine-Gordon action gives the
effective low energy action
    \begin{align}
      \begin{aligned}
	S_{sg} &= 2 \frac{e^2 \epsilon_m}{8 \pi^2} \int d^3\vec r
	\left[ \left( \partial_\mu \chi_1 \right)^2 -
	\frac{M^4}{32\Lambda^2} \cos\left( 2\chi_1 \right) \right]
      \end{aligned}
    \end{align}
after the phase shift $\chi_1 \rightarrow \chi_1 + \pi/2$. This is
also a Sine-Gordon model and leads to the same long distance instanton
physics and confinement as found for even integers. On the other hand,
there is a microscopic difference as this model with odd spins breaks
the lattice symmetry, leading to a two-fold degenerate ground
state. Both models, besides the usual triplon excitations once
confinement is introduced, will also have a massive spin zero mode,
$\chi$ in the first model and $\chi_1$ in the latter.

We now move to the case of odd half integral spin. In this case the
lattice symmetry is broken and the ground state is four-fold
degenerate. The cosine term in the original Sine-Gordon theory has a
phase shift of zero for $\zeta_s$ and $\chi_s$ in the dual $W$
sublattice, a phase shift of $\pi/2$ for the fields in the dual $X$
sublattice, $\pi$ for the fields in the dual $Y$ sublattice, and
$3\pi/2$ for the fields in the dual $Z$ sublattice. Actually these
phase shifts are for a spin $1/2$. For spin $3/2$ the phase shifts are
interchanged among the four sublattices, but lead to similar physics
by relabeling of the sublattices (basically interchanging $X$ and
$Z$). Exactly similar four-fold degeneracies happen for all half-odd
integer spins. The action in this case is
    \begin{widetext}
      \begin{align}
	\begin{aligned}
	  S_{sg} &= \frac{e^2 \epsilon_m a^2}{8 \pi^2} \int_0^{\beta}
	  \frac{d\tau}{a^2} \sum_{\langle s,t \rangle} \left\{ \left(
	  \chi_s^X - \chi_t^Y \right)^2 + \left( \chi_s^X - \chi_t^W
	  \right)^2 + \left( \chi_s^Y - \chi_t^Z \right)^2 + \left(
	  \chi_s^W - \chi_t^Z \right)^2 \right\} \\
	  &+ \frac{e^2 \epsilon_m a^2}{8 \pi^2} \int_0^{\beta} d\tau
	  \sum_t \left[ \left(\frac{\partial \chi_t^W}{\partial \tau}
	  \right)^2 + \left(\frac{\partial \chi_t^X}{\partial \tau}
	  \right)^2 - M^2 \cos \chi_t^W + M^2 \sin \chi_t^X \right] \\
	  &+ \frac{e^2 \epsilon_m a^2}{8 \pi^2} \int_0^{\beta} d\tau
	  \sum_t \left[ \left(\frac{\partial \chi_t^Y}{\partial \tau}
	  \right)^2 + \left(\frac{\partial \chi_t^Z}{\partial \tau}
	  \right)^2 + M^2 \cos \chi_t^Y - M^2 \sin \chi_t^Z \right]
	\end{aligned}
      \end{align}
Defining
      \begin{align}
	\begin{aligned}
	  \chi_W &= \chi_1 + \chi_2 + \chi_3 \;, \qquad \chi_X =
	  \chi_1 - \chi_2 + \chi_4 \\
	  \chi_Y &= \chi_1 + \chi_2 - \chi_3 \;, \qquad \chi_Z =
	  \chi_1 - \chi_2 - \chi_4
	\end{aligned}
      \end{align}
we obtain
      \begin{align}
	\begin{aligned}
	  S_{sg} &= 4\frac{e^2 \epsilon_m}{8 \pi^2} \int d^3r \left\{
	  \left( \nabla \chi_1 \right)^2 + 8 \Lambda^2 \chi_2^2 + 2
	  \Lambda^2 \chi_3^2 + 2 \Lambda^2 \chi_4^2 - 2 \Lambda \chi_1
	  \left( \nabla \chi_4 \right) \right. \\
	  &+ \left. \left( \frac{\partial \chi_1}{\partial \tau}
	  \right)^2 + \left( \frac{\partial \chi_2}{\partial \tau}
	  \right)^2 + \frac{1}{2} \left( \frac{\partial
	  \chi_3}{\partial \tau} \right)^2 + \frac{1}{2} \left(
	  \frac{\partial \chi_4}{\partial \tau} \right)^2 +
	  \frac{1}{2} M^2 \left[ \sin\left( \chi_1 + \chi_2
	  \right)\sin \chi_3 + \cos\left( \chi_1 - \chi_2 \right) \sin
	  \chi_4 \right] \right\} \;.
	\end{aligned}
      \end{align}
    \end{widetext}
Since $\chi_2, \chi_3$ and $\chi_4$ have masses of the order of the
cutoff $\Lambda$, all gradients and time derivatives of $\chi_2,
\chi_3$ and $\chi_4$ are suppressed to zero. We now minimize the
action with respect to $\chi_3$ and $\chi_4$ to get
    \begin{align}
      \begin{aligned}
	\chi_3 &= -\frac{M^2}{8 \Lambda^2} \sin\left( \chi_1 + \chi_2
	\right) \cos\chi_3 \\
	&\simeq -\frac{M^2}{8 \Lambda^2} \sin\left( \chi_1 + \chi_2
	\right) + \mathcal O \left( \frac{M^6}{\Lambda^6} \right) \\
	\chi_4 &= -\frac{M^2}{8 \Lambda^2} \cos\left( \chi_1 - \chi_2
	\right) \cos\chi_4 \\
	&\simeq -\frac{M^2}{8 \Lambda^2} \cos\left( \chi_1 - \chi_2
	\right) + \mathcal O \left( \frac{M^6}{\Lambda^6} \right) \;.
      \end{aligned}
    \end{align}
Substitution of these expressions in the action yields
    \beq
      \nonumber
      S_{sg} = 4\frac{e^2 \epsilon_m}{8 \pi^2} \int d^3r \left\{
      \left( \partial_\mu \chi_1 \right)^2 + 8 \Lambda^2 \chi_2^2
      \right.
    \eneq
    \beq
      - \left. \frac{M^4}{32\Lambda^4} \sin \left( 2 \chi_1\right)
      \sin \left( 2 \chi_2 \right) \right\} \;.
    \eneq
We now minimize with respect to $\chi_2$ to obtain
    \begin{align}
      \begin{aligned}
	\chi_2 &= \frac{M^4}{256 \Lambda^6} \sin \left( 2\chi_1 \right)
	\cos \left( 2\chi_2 \right) \\
	&\simeq \frac{M^4}{256 \Lambda^6} \sin \left( 2\chi_1 \right)
	+ \mathcal O\left( \frac{M^{12}}{\Lambda^{12}} \right) \;.
      \end{aligned}
    \end{align}
Substituting into the action we finally get
    \beq
      S_{sg} = 4\frac{e^2 \epsilon_m}{8 \pi^2} \! \int \! d^3r \!
      \left\{ \left( \partial_\mu \chi_1 \right)^2 -
      \frac{M^8}{8192\Lambda^6} \cos\left( 4 \chi_1\right) \right\}
    \eneq

To summarize, we see that in general instanton effects lead to a
Sine-Gordon action of the form
    \begin{align}
      \begin{aligned}
	&S_{sg} = \frac{e^2 \epsilon_m f(S)}{8 \pi^2} \int d^3r
	\left\{ \left( \partial_\mu \chi \right)^2 \right. \\
	&- \left. \frac{M^2}{2^{\theta\left( f(S) - 3/2 \right)}}
	\left( \frac{M^2}{16 \Lambda^2} \right)^{f(S) - 1} \cos \left(
	f(S) \chi \right) \right\}
      \end{aligned}
    \end{align}
where 
    \beq
      f(S) = 
      \begin{cases}
	1& \quad \text{for even integer } S \\
	2& \quad \text{for odd integer } S \\
	4& \quad \text{for half odd integer } S
      \end{cases} \;.
    \eneq
Finally, if we make the shift $\chi \rightarrow \chi \sqrt{4 \pi^2 /
\left[ e^2 \epsilon_m f(S) \right]}$, we obtain
    \beq 
      \label{finalact} 
      S_{sg} = \int d^3r \left\{ \frac{1}{2} \left(\partial_\mu \chi
      \right)^2 - M^2(S) \cos \left( \chi h(S) \right) \right\}
    \eneq 
with 
    \begin{align} 
      \begin{aligned} 
        h(S) &= \sqrt{\frac{4 \pi^2 f(S)}{e^2 \epsilon_m}} \\ 
	M^2(S) &= \frac{2 \left( ma \right)^{2\rho}}{a^3} \, f(S) \,
        2^{-\theta\left( f(S) - 3/2 \right)} \\
	&\times \left(\frac{\pi^2 \left( ma \right)^{2\rho}}{e^2
        \epsilon_m a} \right)^{f(S) - 1} \;.
    \end{aligned} 
  \end{align}

We see that the effects of instantons are described by a Sine-Gordon
theory. The conclusion that instanton effects can be described by a
Sine-Gordon theory is true for all microscopic spins. The only
difference is in the mass of the Sine-Gordon theory we just wrote and
in the factor inside the cosine of the Sine-Gordon theory. Hence the
long distance confinement consequences of tunneling effects are
equivalent independent of the microscopic spins with just irrelevant
numerical differences. Even though we did not discuss it in detail
here, the Berry phases do make a difference for the paramagnetic
ground state, which leads to breaking of lattice
symmetries\cite{hal2,sachread}. They make the ground state quadruply
degenerate for half-odd integer spins, doubly degenerate for odd
integer spins, and a non-degenerate valence bond solid for even
integer spins. {\it But as far as the instantons the long distance
confinement physics is the same regardless of microscopic spins.}

\subsection{Confinement of $z$ Spinons in the Quantum Paramagnetic Phase}

In order to see that the effects of instanton fluctuations lead to
confinement, we now closely follow Polyakov and calculate the
correlation or Green's function between electromagnetic fields. The
electromagnetic field is defined as before
    \beq
      H_\mu (x) = \frac{1}{2} \epsilon_{\mu\nu\lambda} F_{\nu \lambda}
      (x) \,.
    \eneq
The instantons, or the instanton charge density, acts as a source for
these electromagnetic fields via
    \beq
      \label{instemf0}
      H_\mu (x) = \frac{1}{2} \int d^3y \frac{(x - y)_\mu}{|\vec x -
      \vec y|^3} \rho (\vec y)
    \eneq
or in momentum space
    \beq
      H_\mu (k) = 2\pi i \frac{k_\mu}{k^2} \rho(k) \,.
    \eneq

It proves very convenient to introduce sources for our instantons. In
particular we calculate
    \beq
      \label{expesp}
      \langle e^{i\int \rho(x)\eta(x) d^3x} \rangle =
      \frac{Z[\eta(x)]}{Z[0]} \,,
    \eneq
where
    \begin{align}
      \begin{aligned}
	\label{rhoZ}
	\rho(x) &= \sum_a q_a \delta(x - x_a) \\
	Z_{inst} [\eta] &= \int \mathcal D\chi \exp\left\{-\int d^3r
        \left[ \frac{1}{2} \left[ \partial_\mu \left( \chi - \eta
        \right) \right]^2 \right. \right. \\
	& \qquad -\left. \left. M^2(S) \cos \left( \chi h(S) \right)
        \right] \right\} \;.
      \end{aligned}
    \end{align}
Correlation functions of $\rho$ can be derived from the partition
function by taking derivatives with respect to $\eta$: $\partial^{(n)}
Z_{inst} / \partial \eta^{(n)}$. In particular, we obtain
    \beq
      \langle \rho(k) \rho(-k) \rangle = k^2 - k^4 \langle \chi(k)
      \chi(-k) \rangle
    \eneq

The correlation function for the electromagnetic fields is given by
    \begin{align}
      \begin{aligned}
	\langle H_\mu(k) H_\nu(-k) \rangle &= \langle H_\mu (k)
	H_\nu(-k) \rangle^{(0)} \\
	&+ \frac{k_\mu k_\nu}{k^4} \langle \rho(k) \rho(-k) \rangle
      \end{aligned}
    \end{align}
where $\langle H_\mu(k) H_\nu(-k)\rangle^{(0)}$ is the Green's
function without instantons. With the appropriate screening, this
Green's function is
    \begin{align}
      \begin{aligned}
	\langle H_\mu H_\nu\rangle^{(0)} &= \frac{1}{\epsilon_m k^2}
	\left( k^2 \delta_{\mu\nu} - k_\mu k_\nu \right) \\
	&= \frac{1}{\epsilon_m} \left[ \delta_{\mu\nu} - \frac{k_\mu
	k_\nu}{k^2} \right]
      \end{aligned}
    \end{align}
The $k=0$ pole reflects the masslessness of the photon. Now we
calculate the instanton-instanton density correlation function. Before
doing so we note that in the dilute gas approximation the $\chi$
coupling constant is extremely small ($ma \ll 1$) and thus provides
only a small renormalization without changing the qualitative behavior
of the theory. Hence the Sine-Gordon cosine potential may be
approximated to second order in $\chi$ to high accuracy and
    \beq
      \langle \chi(k) \chi(-k)\rangle \simeq \frac{1}{k^2 + \left(h(S)
      M(S)\right)^2} \,.
    \eneq
We find that the instanton charge density correlator is given by
    \begin{align}
      \begin{aligned}
	\langle \rho(k) \rho(-k) \rangle &= k^2 - \frac{k^4}{\left(
	h(S) M(S)\right)^2 + k^2} \\
	&= \frac{\left( h(S) M(S)\right)^2 k^2}{k^2 + \left( h(S) M(S)
	\right)^2} \;.
      \end{aligned}
    \end{align}
The electromagnetic propagator becomes
    \begin{align}
      \begin{aligned}
	&\langle H_\mu(k) H_\nu(-k) \rangle = \\
	&\frac{1}{\epsilon_m} \left[\delta_{\mu\nu} - \frac{k_\mu
	k_\nu}{k^2} + \frac{k_\mu k_\nu}{k^2} \frac{ \left( h(S)M(S)
	\right)^2}{ k^2 + \left( h(S)M(S) \right)^2 } \right] \\
	&= \frac{1}{\epsilon_m} \left[ \delta_{\mu\nu} - \frac{k_\mu
	k_\nu}{k^2 + \left( h(S) M(S) \right)^2} \right] \\
      \end{aligned}
    \end{align}

We see from the pole in the last term that the photon acquired a mass
$M_p = h(S) M(S)$ without spontaneous breaking of $U(1)$
symmetry. This is a sign of confinement and that the ground state is a
gauge singlet. If the reader is unsatisfied with this correct, but
indirect conclusion, we can check that the theory indeed confines by
calculating that the Wilson loop\cite{wilson, polyaqed, polyab}
    \beq
      F[C] \equiv e^{-\mathcal W[C]} \equiv \langle e^{i\oint A_\mu
      dx_\mu} \rangle \,.
    \eneq
gives an area law\cite{polyaqed,polyab}. Explicitly, using Stokes'
theorem we have
    \beq
      F[C] = \langle e^{i\oint A_\mu dx_\mu} \rangle = \langle
      e^{i\int_A H_\mu dS_\mu} \rangle
    \eneq
which, using equation (\ref{instemf0}) can be written as
    \beq
      F[C] = \langle e^{i\int \eta(x) \rho(x) d^3x} \rangle
    \eneq
with
    \beq
      \eta(x) = \frac{1}{2} \int_A dS_y \cdot \frac{(\vec x - \vec
      y)}{|\vec x - \vec y|^3} \,.
    \eneq
Using (\ref{expesp}) and (\ref{rhoZ}) we find
    \begin{align}
      F[C] &= \frac{1}{Z[0]} \int \mathcal D\chi \exp\left\{-\int d^3r
      \left[ \frac{1}{2} \left[ \partial_\mu \left( \chi - \eta
      \right) \right]^2 \right. \right. \\
      \nonumber
      &- \left. \left. M^2(S) \cos \left( \chi h(S) \right) \right]
      \right\} \;.
    \end{align}
In the saddle point approximation we obtain
    \begin{align}
      \begin{aligned}
	F[C] &\sim \exp\left\{-\int d^3r \left[ \frac{1}{2} \left[
	\partial_\mu \left( \chi_{cl} - \eta \right) \right]^2
	\right. \right. \\
	&- \left. \left. M^2(S) \cos \left( \chi_{cl} h(S) \right)
	\right] \right\} \;.
      \end{aligned}
    \end{align}
The classical field is obtained by solving the equation
    \beq
      \partial^2 (\chi_{cl} - \eta) = M^2 (S) h(S) \sin\left(
      \chi_{cl} h(S) \right)
    \eneq
which takes the form
    \begin{align}
      \nonumber
      \partial^2\chi_{cl} &= 2\pi \delta^1(z)\, \theta_A (xy) +
      M^2 (S) h(S) \sin \left( \chi_{cl} h(S) \right) \\
      \theta_A(xy) &= \begin{cases} 1& \qquad x,y\in A \\ 
	0& \qquad \text{otherwise} \end{cases}  \quad .
    \end{align}
This has solution
    \beq
      \chi_{cl} (z) = \begin{cases} 4 \arctan\left( e^{-M(S) h(S)
      z}\right) \;, \qquad z > 0 \\ -4 \arctan\left( e^{M(S) h(S) z}
      \right) \;, \qquad z < 0 \end{cases} \;.
    \eneq
It now immediately follows that 
    \begin{align}
      \begin{aligned}
	F[C] &\simeq e^{-\gamma (S) A} \\
	\gamma (S) &= \frac{1}{2} \int_{-\infty}^{\infty} dz \,
	(\chi_{cl} - \eta)(z) (\chi_{cl}'' - \eta'')(z) \\
	&+ \int_{-\infty}^{\infty} dz \, M^2 (S) \cos \left(
	\chi_{cl}(z) h(S) \right)
      \end{aligned}
    \end{align}
where $A$ is the area of the $xy$-plane. 

We see that the Wilson loop satisfies the area law. The interaction
energy between charges is given by
    \beq
      E(R) = \gamma (S) R
    \eneq
and the theory confines for all values of the microscopic spins. The
force between spinons is constant and equal to the ``string tension''
$\gamma (S)$. The scale below which the theory confines $1/\xi(S) =
M(S) h(S)$ is spin dependent but nonzero. Basically spinons are
confined at length scales larger than $\xi(S)$ and are not relevant to
the physics of the system. The low energy excitations are spin 1
triplons ($z^\dagger \vec \sigma z$) and a spin 0 collective mode
corresponding to the Sine-Gordon field $\chi$.

\section{ Quantum Critical Deconfinement of Spinons }

We saw in the previous section that spinons are confined and that the
confinement arose from the topological tunneling events originating
from the compactness of the theory. On the other hand, it has recently
been suggested that there is a new class of quantum critical points
whose properties are controlled by the deconfinement of spinons rather
than fluctuations of the order parameter\cite{bob1, sachdev2}. In
particular, it was proposed that this new kind of quantum critical
points occur in $2 + 1$ D antiferromagnets. Finally, it was further
proposed that, because of the dependence on microscopic spins of the
Berry phases, the deconfinement occurs only for spin $1/2$
systems\cite{sachdev2}. In the present section, we will study how
deconfinement occurs and under which conditions.

The quantum critical point occurs when the spinons become massless, $m
= 0$. As we saw above, the only dependence on the microscopic spins,
as far as long distance confinement properties, appears in the
parameter of the Sine-Gordon theory that describes the nontrivial
tunneling effects in $2+1$ D antiferromagnets. As shown in the
previous section, when $m > 0$ antiferromagnets confine spinons for
all microscopic spins. The presence of confinement is characterized by
the nonzero value of the photon mass or confinement scale
    \begin{widetext}
      \beq
        M_p(S) = M(S) h(S) = \sqrt{\frac{8 \pi^2 \left( ma
	\right)^{2\rho}}{e^2 \epsilon_m a^3} \, f^2(S) \,
	2^{-\theta\left( f(S) - 3/2 \right)} \left(\frac{\pi^2 \left(
	ma \right)^{2\rho}}{4 e^2 \epsilon_m a} \right)^{f(S) - 1}}
	\;,
      \eneq
    \end{widetext}
such that confinement occurs for all energies {\it smaller} than $M_p
(S)$. It now follows that the confinement energy scale vanishes at the
quantum critical point for all values of the microscopic spin. {\it
Therefore, spinons are deconfined at the quantum critical point
independent of the microscopic spin}. On the other hand, there is a
dependence of microscopic spins on the approach to the critical
point. For half-odd integer spin systems the confinement length scale
$\xi(S) = 1/M_p(S)$ diverges faster than for odd integer spin
systems. The confinement length scale diverges faster for odd integer
spin systems than for even integer spin systems. Therefore the only
dependence on the microscopic spin is how fast one reaches the
deconfined quantum critical point, but deconfinement occurs for all
microscopic spins.

The conclusion that deconfinement is independent of microscopic spins
is new and unexpected. Therefore we will check that this is indeed so
with more care. At face value the only conclusion that can be reached
is that at $m = 0$, $M(S) = 0$. Let's calculate the Wilson loop for
the case of $M(S) \rightarrow 0$ and see if the theory confines or
not. As we saw in the previous section, to calculate the Wilson loop
we need to evaluate
    \begin{align}
      \begin{aligned}
	F[C] &\sim \exp\left\{-\int d^3r \left[ \frac{1}{2} \left[
	\partial_\mu \left( \chi_{cl} - \eta \right) \right]^2
	\right. \right. \\
	&- \left. \left. M^2(S) \cos \left( \chi_{cl} h(S) \right)
	\right] \right\} \;.
      \end{aligned}
    \end{align}
where 
    \beq 
      \partial^2 (\chi_{cl} - \eta) = M^2 (S) h(S) \sin\left(
      \chi_{cl} h(S) \right) \;.
    \eneq 
When $m = 0$, $M(S) = 0$ and we have 
    \beq 
      F[C] \sim \exp\left\{- \frac{1}{2} \int d^3r \left[ \partial_\mu
      \left( \chi_{cl} - \eta \right) \right]^2 \right\}
    \eneq 
with 
    \beq 
      \partial^2 (\chi_{cl} - \eta) = 0 \;.
    \eneq 
Now the Wilson loop is calculated very straightforwardly by
integrating by parts to obtain
    \beq 
      F[C] \sim \exp\left\{ \frac{1}{2} \int d^3r \left( \chi_{cl} -
      \eta \right) \partial^2 \left( \chi_{cl} - \eta \right) \right\}
      = 1 \;.
    \eneq
Since $F[C] = e^{-ER}$, the long distance force between spinons is
zero and they are deconfined. This result is independent of
microscopic spins as $M(S) \rightarrow 0$ for all microscopic spins.

There is one last but equivalent way to see that deconfinement occurs
independent of microscopic spin, i.e. that the instantons or
compactness effects disappear at the quantum critical point. If we go
back to the partition function from which all the confinement physics
followed
    \begin{align}
      \begin{aligned}
	\mathcal Z &= \int \mathcal D z \mathcal D z^\dagger \mathcal D
        A_\mu e^{-S_{zA}} \, \mathcal Z_{inst} \\
	S_{zA} &= \frac{1}{4 e^2 \epsilon_m} \int d^3r F_{\mu\nu}^2 \\
	\mathcal Z_{inst} &= \sum_{K, q_s} \frac{\zeta^K}{K!}
        \prod_{s=1}^K \left[ \sum_{R_a} \int_0^{\beta}
        \frac{d\tau_s}{a} \right] e^{-S_{inst}} \\
        S_{inst} &= \frac{\pi}{2 e^2 \epsilon_m} \sum_{s\neq t}
        \frac{q_s q_t}{\left[ (\vec R_s - \vec R_t)^2 + (\tilde \tau_s
        - \tilde \tau_t)^2 \right]^{1/2}} \\
	&+ i\sum_s \pi S \zeta_s q_s \\
	\zeta &= e^{-S_{cl}} = \left[ \frac{m}{\Lambda} \right]^{2
	\rho (q)} \;.
      \end{aligned}
    \end{align}
$\zeta$ is the instanton fugacity as calculated before\cite{murthsach}
and $\rho (q) > 0$ except when we have no instantons, where by
definition $\rho (0) \equiv 0$. $\zeta = 1$ for this last case. It now
follows immediately that at the critical point $\zeta = 0$ except in
the absence of instantons, $K = 0, q = 0$. Therefore, $\mathcal
Z_{inst} = 1$, there are no Berry phase terms and no monopole events
and the theory is deconfined independent of microscopic spins.

\section{Spinon Deconfined Critical Theory}

As we obtained in the previous section, at the quantum critical point,
when the spinon mass $m$ vanishes, the instanton and Berry phase terms
disappear from the theory and the quantum critical point corresponds
to massless spinons coupled to an emergent $U(1)$ gauge field with no
effects of topology. The emergent gauge group is noncompact at
criticality. We had calculated before the effects of the spinons on
the emergent photon and found that when the spinons were integrated
out, we obtain ``dielectric screening'' of the emergent
electromagnetic fields. This is easily understood as the spinons have
a gap and hence the theory is a ``semiconductor'' as far as the
emergent electromagnetic properties. The spinons are ``semimetallic''
at criticality and we need to calculate the screening properties in
this case anew.

The critical theory is described by the action
    \begin{align}
      \begin{aligned}
	S &= \frac{2}{ga} \int d^3 r \, |(\partial_\mu - i A_\mu)z|^2
	\\
	&+ \int d^3 r \left\{ i\delta\lambda (|z|^2 - 1) +
	\frac{1}{4e^2} F_{\mu\nu}^2 \right\} \;.
      \end{aligned}
    \end{align}

As before, $\delta\lambda$ is a Lagrange multiplier to enforce the
constraint $|z|^2 = 1$. Just as the gauge fields $A_\mu$ acquire
dynamics, spinon fluctuations give dynamics to the Lagrange multiplier
$\delta\lambda$. The dynamics for $\delta\lambda$ follows from the
same one loop spinon fluctuations that gave dynamics to the gauge
fields. To one loop order 
    \beq 
      \langle \delta\lambda(k) \delta\lambda(-k) \rangle \simeq
      \frac{1}{\pi^2} \left( \frac{g}{2\Lambda} \right)^2 \frac{1}{k}
      \;.
    \eneq
This term is the same as would be obtained from expanding the
exponential of the action
    \beq
      \frac{1}{\pi^2} \left( \frac{g}{2\Lambda} \right)^2 \int
      \frac{d^3k}{(2\pi)^3} \frac{1}{k} \, \delta \lambda(k) \,
      \delta\lambda(-k) \;.
    \eneq
The $1/k$ means that $\delta\lambda$ fluctuations will be suppressed
at long wavelengths and hence irrelevant to the low energy
physics. For completeness, we keep these terms in the effective
action, which leads to a $\delta\lambda$ propagator
    \beq
      \pi^2 \left( \frac{2\Lambda}{g} \right)^2 k \;.
    \eneq

Now we must calculate the screening effect of the massless
spinons. This is easily done via the renormalization group by
integrating high energy spinon degrees of freedom and computing their
effects on the renormalization of the electric charge. In this way we
obtain the beta function
    \beq
      \tilde \beta_{e^2} = \mu \frac{\partial}{\partial_\mu} e_\mu^2 =
      \frac{e_\mu^4}{\pi^2 \mu} \,.
    \eneq
This function is not dimensionless as usual because in $2+1$
dimensions the electric charge has dimension of square root of mass or
momentum. This equation can be integrated easily and yields the
renormalized charge at scale $\mu \ll \Lambda$
    \beq
      e_\mu^2 \simeq \pi^2 \mu \,.
    \eneq
Therefore, from $e_\mu^2 = e^2/\epsilon_m$, we obtain the screening
factor at scale $\mu = k$ in momentum space
    \beq
      \epsilon_m = \frac{e^2}{\pi^2 k} \,.
    \eneq

Spinons interact through photon exchange. The effective interaction
between spinons is given by the photon-photon Green's function, which
we calculate in the Lorentz gauge
   \beq
      V_{\mu\nu} = \langle A_\mu (k) A_\nu (-k) \rangle =
      \frac{\pi^2}{e^2 k} \left[ \delta_{\mu\nu} - \frac{k_\mu
      k_\nu}{k^2} \right] \,.
    \eneq
which in real space is formally written as
    \beq
      V_{\mu\nu} (x - x') = \frac{\pi^2}{e^2 (x - x')^2} \left[
      \delta_{\mu\nu} - \partial_\mu \frac{1}{\partial^2} \partial_\nu
      \right] \,.
    \eneq
Therefore, the effective theory at criticality is described by the
action
    \begin{align}
      \begin{aligned}
	S &= \int d^3r \left[ \frac{2\Lambda}{g} |\partial_\mu z|^2 +
	i \delta\lambda (|z|^2 - 1) \right] \\ 
	&- \frac{1}{2} \left(\frac{2\Lambda e}{g}\right)^2 \int d^3 r
	d^3r' J_\mu (r) V_{\mu\nu} (r - r') J_\nu (r') \\
	&+ \int d^3r J^a(r) z^\dagger (r) \sigma^a z (r)
      \end{aligned}
    \end{align}
where
    \beq
      J_\mu (r) = i \left( z^\dagger \partial_\mu z - z
      \partial_\mu z^\dagger \right) \,.
    \eneq
We have added an external source $J^a(r)$ that couples to the
N\'eel field $n^a = z^\dagger \sigma^a z$ because we want to study the
$\langle n^a(r) n^b(r') \rangle$ correlator. 

In order to analyze the critical action, we introduce a
Hubbard-Stratonovich real vector field $B_\mu$ to write an action
linear rather than quadratic in $J_\mu$. The action for the nonlinear
sigma model at criticality is then
    \begin{align}
      \nonumber
      S &= \int d^3r \left[ \frac{2\Lambda}{g} |\partial_\mu z|^2 + i
      \delta\lambda (|z|^2 - 1) \right] \\
      \label{seffB}
      &+ \frac{1}{2} \int d^3 r d^3r' B_\mu (r) \tilde V_{\mu\nu}^{-1}
      (r - r') B_\nu (r') \\
      \nonumber
      &+ \frac{2 \Lambda e}{g} \int d^3 r B_\mu (r) J_\nu (r) + \int
      d^3r J^a(r) z^\dagger (r) \sigma^a z (r) \;.
    \end{align}
with
    \beq
      \tilde V_{\mu\nu}^{-1} (k) = \frac{e^2 k}{\pi^2} \left[
      \delta_{\mu\nu} - \frac{k_\mu k_\nu}{k^2} \right]
    \eneq
Of course, in the partition function we must integrate over $z,
z^\dagger, \delta\lambda$ and $B_\mu$. The $B_\mu$'s are gauge fields
with the appropriate gauge invariance as can be seen from their
action. We can now integrate the spinon fields $z$ to obtain the
partition function 
    \beq 
      Z = \int \mathcal D B_\mu \mathcal D \delta\lambda \;
      e^{-S_{\text{eff}}}
    \eneq 
where $S_{\text{eff}}$ is, in momentum space for convenience
    \begin{align}
      \begin{aligned}
	S_{\text{eff}} &= \exp\left( - \int \frac{d^3k}{(2\pi)^3}
	\left[ \frac{1}{2} B_\mu (-k) \tilde V_{\mu\nu}^{-1} (k) B_\nu
	(k) - i \delta\lambda \right] \right. \\
	&- \left. \text{Tr} \ln M \right) \;
      \end{aligned}
    \end{align}
The operator $M$ is 
    \begin{align}
      \begin{aligned}
	&M (q_1,q_2) = \left( i \frac{g}{2\Lambda} \delta\lambda +
	q_1^2 \right) \; \delta (q_1 - q_2) \\
	&+ B_\mu (-q_1 + q_2) \left[\frac{q_1 + q_2}{2} \right]_\nu \\
	&+ B_\mu (q_1 - q_2) \left[ \frac{q_1 + q_2}{2} \right]_\nu +
	J^a (q_1 - q_2) \sigma^a \;. \\
      \end{aligned}
    \end{align}
We can now move to calculate the N\'eel field correlator
    \begin{align}
      \begin{aligned}
	&\langle n^a(k) n^b (-k)\rangle \delta (k + q) \equiv \langle
	n^a (k) n^b (q) \rangle \\
	&= \frac{1}{Z} \frac{\partial^2 Z}{\partial J^a(k) \partial
	J^b (q)} \\
	&= \frac{1}{Z} \text{Tr} \left( \sigma^a \sigma^b \right) \int
	\mathcal D B_\mu \mathcal D \delta\lambda \int \frac{d^3 q_1
	d^3 p}{(2\pi)^6} \\
	&\times M^{-1} (q_1, p) M^{-1} (p-k, q+q_1) \Big |_{J=0} \;
	e^{-S_\text{eff}} \\
	&= \frac{2\, \delta^{ab}}{Z} \int \mathcal D B_\mu \mathcal D
	\delta\lambda \int \frac{d^3 q_1 d^3 p}{(2\pi)^6} \\
	&\times M^{-1} (q_1, p) M^{-1} (p-k, q+q_1) \Big |_{J=0} \;
	e^{-S_\text{eff}} \,.
      \end{aligned}
    \end{align}
To lowest order we find
    \begin{align}
      \begin{aligned}
	&\langle n^a(k) n^b(-k)\rangle = 2 \delta^{ab} \int
	\frac{d^3p}{(2\pi)^3} \frac{1}{p^2 (p-k)^2} \\
	&= \frac{\delta^{ab}}{k} \left( \frac{1}{8} + \frac{1}{\pi^2}
	\right) \simeq \frac{\delta^{ab}}{k} 0.25 \;.
      \end{aligned}
    \end{align}
We point out that this is what is obtained from non-interacting, free
spinons. That is, {\it the N\'eel field decays into two free spinons,
which are then reabsorbed to reconstitute the N\'eel field}. We see
then that {\it the N\'eel field Green's function has anomalous
exponent $\eta = 1$. This is the unique consequence of decay of the
N\'eel magnetization into free spinons at criticality}. In our model,
the spinons are expected to be free as they can interact through
emergent photons, but at the critical point the electromagnetic
potentials acting between the charge currents generated by the spinons
are screened from the Coulomb form $1/k^2$ to $1/k$, leading to a
retarded interaction $\sim 1/ (\vec R^2 + \tau^2)$, and to a Coulomb
law $\sim 1/|\vec R|$. Hence the spinons are expected to be
free. Let's see if this is indeed so. The one loop correction coming
from the effective spinon interaction is given by
    \begin{align}
      \nonumber
      &\langle n^a (k) n^b(-k)\rangle \Big|_{1\text{ loop}} \\
      \nonumber
      &= 8 \pi^2 \delta^{ab} \int \frac{d^3 q_2 d^3p}{(2\pi)^6}
      \frac{p\cdot (k+p)}{q_2 (p - q_2)^2 p^2 (p + k)^2 (k + p -
      q_2)^2} \\
      &\simeq \frac{\delta^{ab}}{k} 0.104329 \,.
    \end{align}
Therefore, including the interactions to first order we find that the
spinons still behave as free as the anomalous exponent $\eta=1$. The
only effect of the interactions was to renormalize the non-universal
constant of proportionality multiplying $1/k$. 

Our last result is a one photon interaction result, but a simple
exercise in Feynman diagrams using the Feynman rules that follow from
the effective action (\ref{seffB}) shows that higher order
corrections lead to corrections proportional also to $1/k$. Inclusion
of Lagrange multiplier $\delta \lambda$ lines or photon lines internal
to the spinon bubble
    \begin{figure}[!h]
      \includegraphics[width=3.8cm, height=0.8cm]{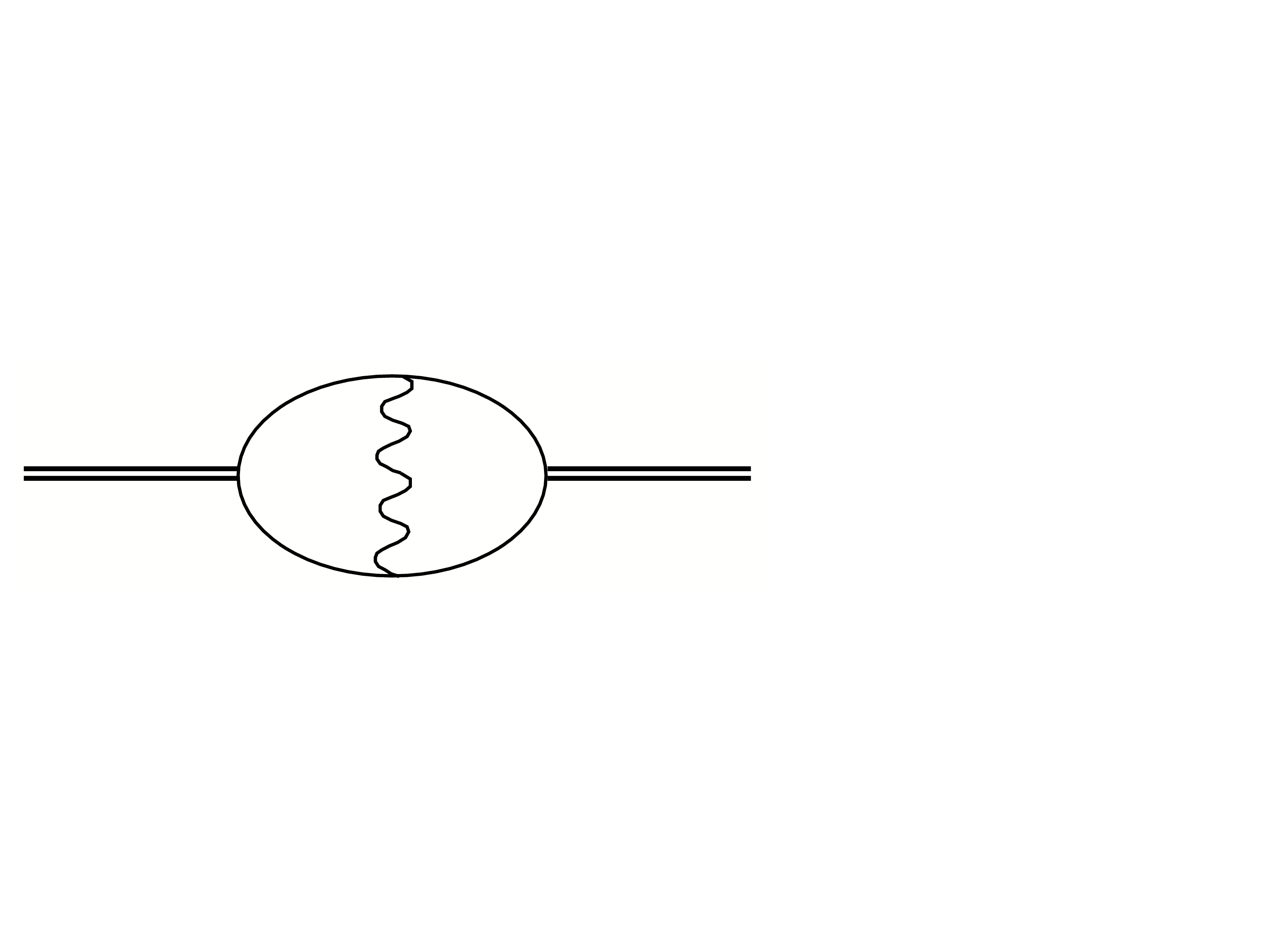}
      \includegraphics[width=3.8cm, height=0.85cm]{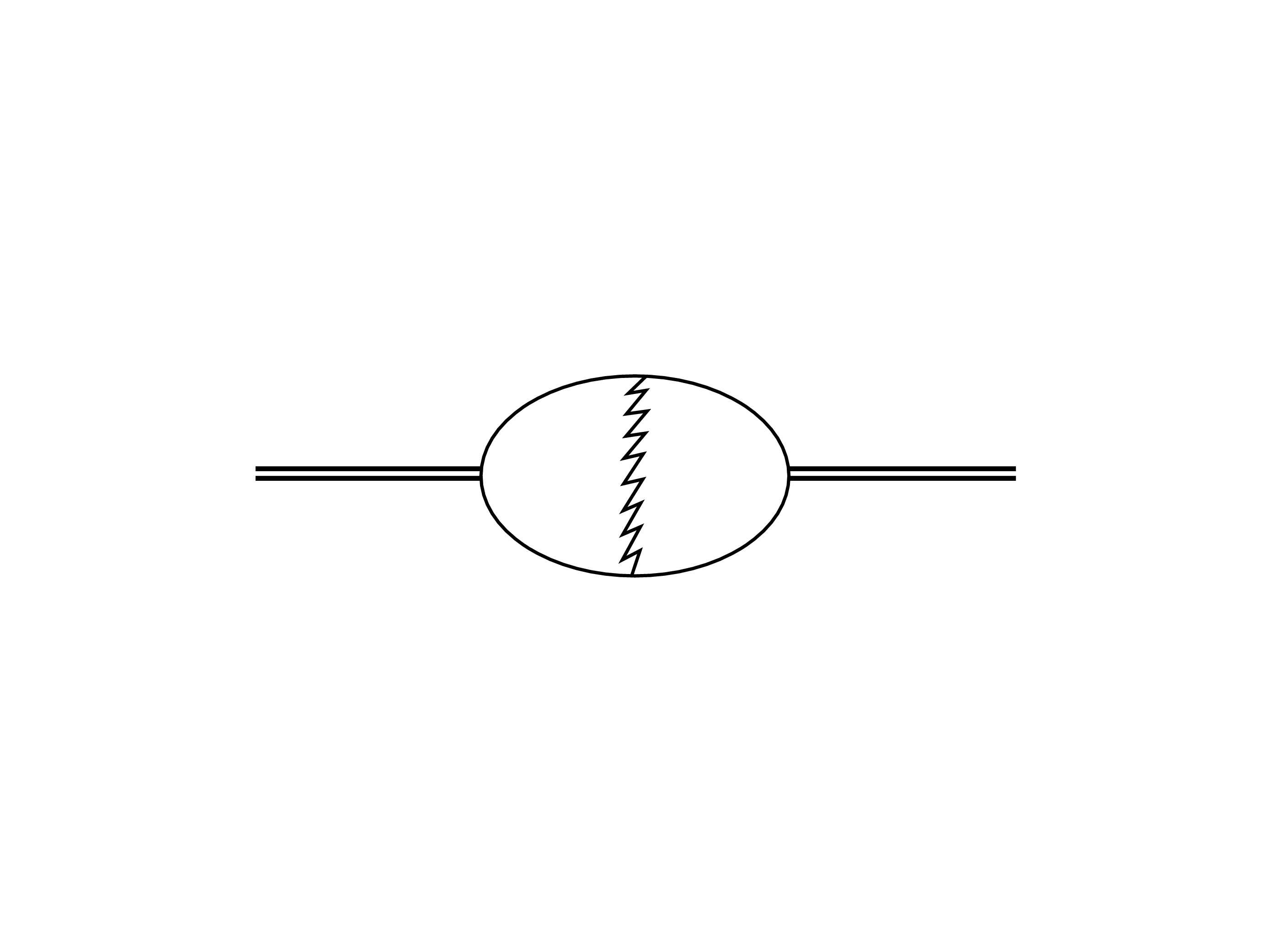}
    \end{figure}

\noindent gives contributions proportional to $1/k$ by direct
computation or by simple power counting. In these diagrams, the double
solid lines correspond to N\`eel field $n$ lines, the single solid
lines corresponds to spinon $z$ or $z^\dagger$ lines, the wiggly line
corresponds to a photon propagator and the lightning bolt corresponds
to $\delta \lambda$ lines or propagators. Similarly, inclusion of
further internal lines to the bubble gives contributions proportional
to $1/k$. There are also corrections given by the diagrams
    \begin{figure}[!h]
      \includegraphics[width=3.8cm, height=0.75cm]{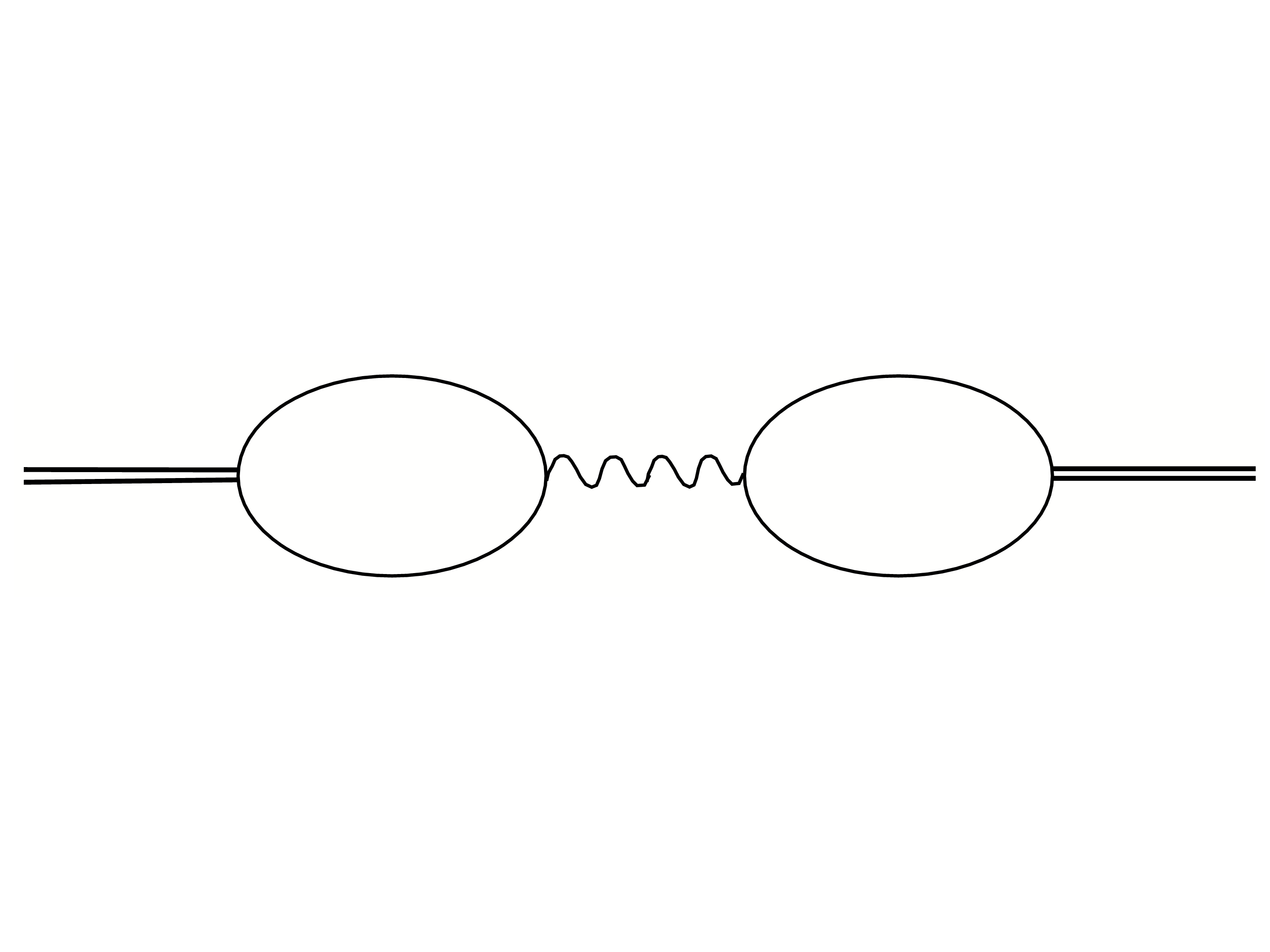}
      \includegraphics[width=3.8cm, height=0.8cm]{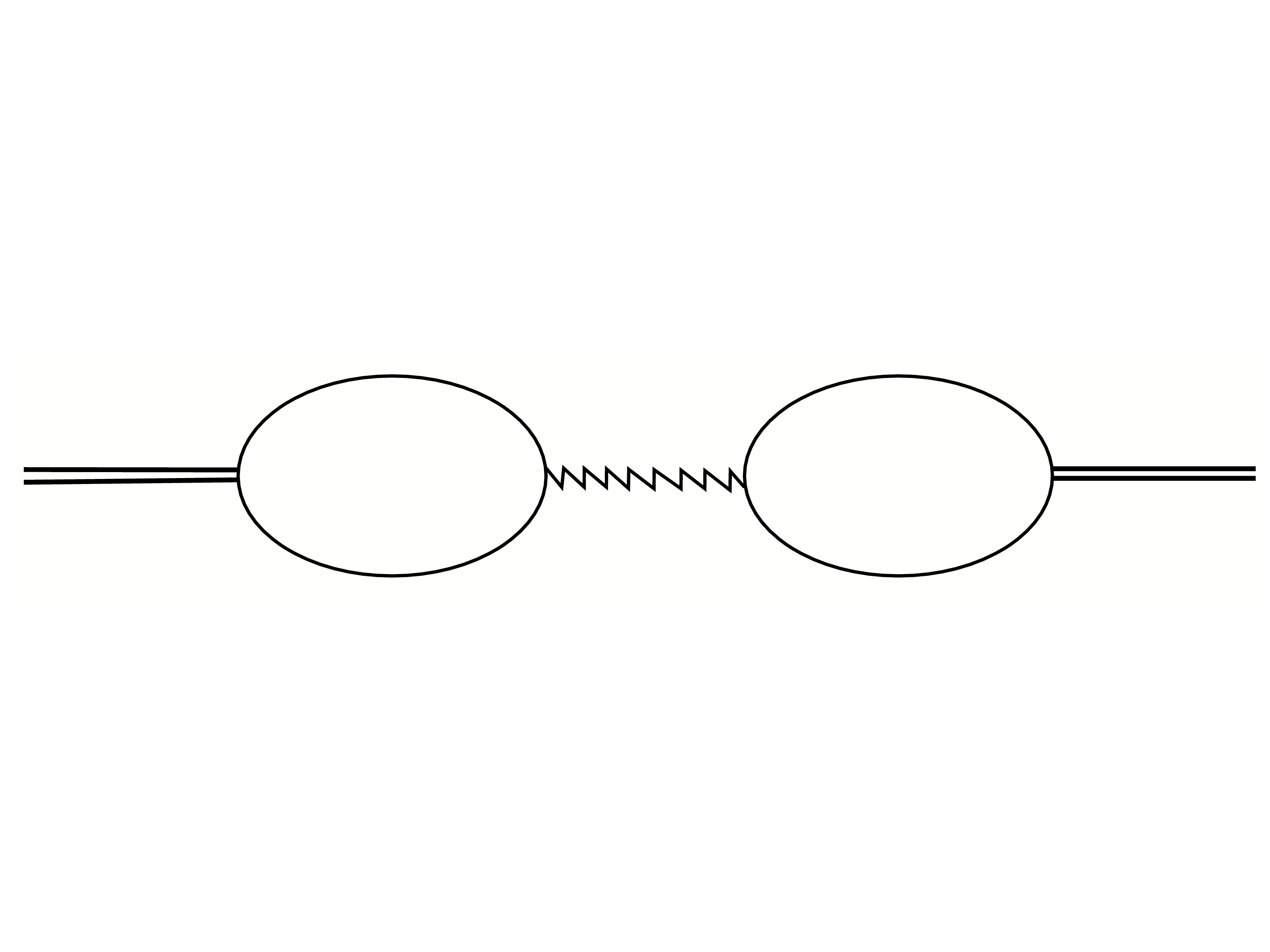}
    \end{figure}

\noindent which give contributions proportional to $1/k$. Inclusion of
higher order bubbles or corrections internal to the spinon bubble also
yield contributions proportional to $1/k$. One nontrivial point which
could invalidate the conclusion that $\eta=1$ is if internal self
energy corrections to the spinon propagator include $\ln k$
terms. Explicit calculation of the spinon self energy corrections
shows that the log terms that are generated are multiplied by factors
of $k^2$ which make them vanish as $k \rightarrow 0$ and are thus
irrelevant to the universal low energy critical physics. Hence we see
that {\it at criticality, spinons are not only deconfined but behave
as noninteracting at long distances and lead to critical exponent
$\eta = 1$ equivalent to that of decay of the N\`eel field into
critical free spinons}.

The deconfined critical points studied and elucidated here seem to be
different from the $2 + 1$ D Heisenberg critical points. It has been
suggested before that these two different types of critical points
might occur in $2+1$ D\cite{sachdev2}. One particular suggestion is
that interactions irrelevant to the N\`eel and quantum paramagnetic
phases turn the Heisenberg critical point into a deconfined critical
point and there seems to be indirect numerical evidence for such
physics\cite{sandvik}. In references\cite{sandvik}, evidence for a
continuous transition between a valence bond ordered paramagnet and
its corresponding N\'eel ordered phase was presented in both
Heisenberg and $XY$ systems. Evidence for a relatively large value of
$\eta$ was also presented. It has been suggested that these
transitions correspond to the deconfined critical
points\cite{sachdev2} studied in the present work.

We have shown that if a deconfined critical point exists in $2+1$ D
antiferromagnets, it will occur irrespective of the microscopic spin
of the system, and that the confinement length will diverge faster
upon approach to criticality for half-odd integer spins, next fastest
for odd integer spins, and slowest for even integer spins as a
consequence of the Berry phase terms relevant to the quantum
paramagnetic phase. Deconfinement occurs because instanton or monopole
events vanish at criticality\cite{murthsach} and together with them,
the Berry phase terms vanish too\cite{hal2}. We also find that the
emergent photon at criticality is screened strongly at long distances
making the spinons behave as if they were strictly free at long
wavelengths. Finally, the N\`eel critical correlator behavior follows
from decay into free spinons, which universally leads to the critical
exponent $\eta=1$. This is a diagnostic of deconfined criticality,
that is, {\it for a free deconfined spinon critical point we predict a
critical exponent $\eta$ exactly equal to one}.

\section{Some Experimental Consequences of Critical Spinon Deconfinement}

Now that we have found the effective critical theory of deconfined
critical points, we will briefly obtain some experimental consequences
of the free spinon critical theory. As we have seen, the N\'eel
critical propagator or susceptibility is given by 
    \begin{align}
      \begin{aligned}
	&\langle n_a(\omega,\vec k) n_b(-\omega,-\vec k) \rangle =
	G_{ab} (\omega, \vec k) \\
	&\equiv \delta_{ab} G (\omega, \vec k) = \delta_{ab} \,
	\frac{C}{\Lambda} \frac{1}{\sqrt{\vec k^2 - \omega^2}}
      \end{aligned}
    \end{align}
where $C$ is a dimensionless constant. Experimental measurable
quantities are usually proportional to the density of states, which at
finite temperatures is given by
    \beq 
      D (\omega, \vec k, T) = \frac{1}{\pi} \, \text{Im} \, G (\omega,
      \vec k) \, \frac{1}{1 - e^{-\omega / T}} \;.  \eneq The
      imaginary part is easily calculated to be \beq \text{Im} \, G
      (\omega, \vec k) = \frac{\theta \left( |\omega| - |\vec k|
      \right)}{\sqrt{\omega^2 - \vec k^2}}
    \eneq 
to give a density of states
    \beq 
      \nonumber
      D (\omega, \vec k, T) = \frac{1}{\pi} \theta \left( |\omega| -
      |\vec k| \right) \frac{1}{\sqrt{\omega^2 - \vec k^2}} \,
      \frac{1}{\left(1 - e^{-\omega / T}\right)} \;.
    \eneq

    \begin{figure}
      \begin{center}
        \subfigure{\includegraphics[width=5.5cm,
        height=4cm]{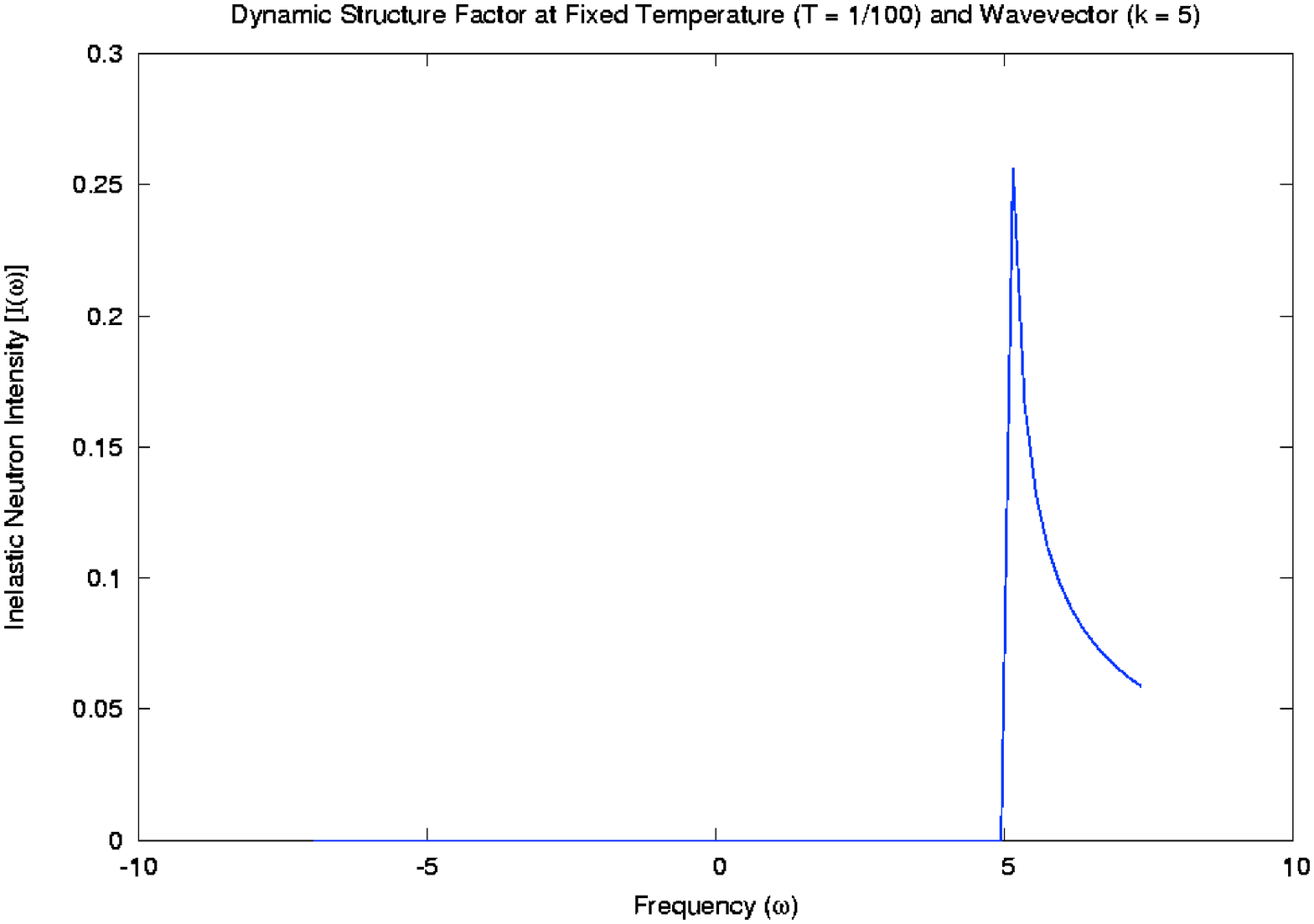}}
	\subfigure{\includegraphics[width=5.5cm,
	    height=4cm]{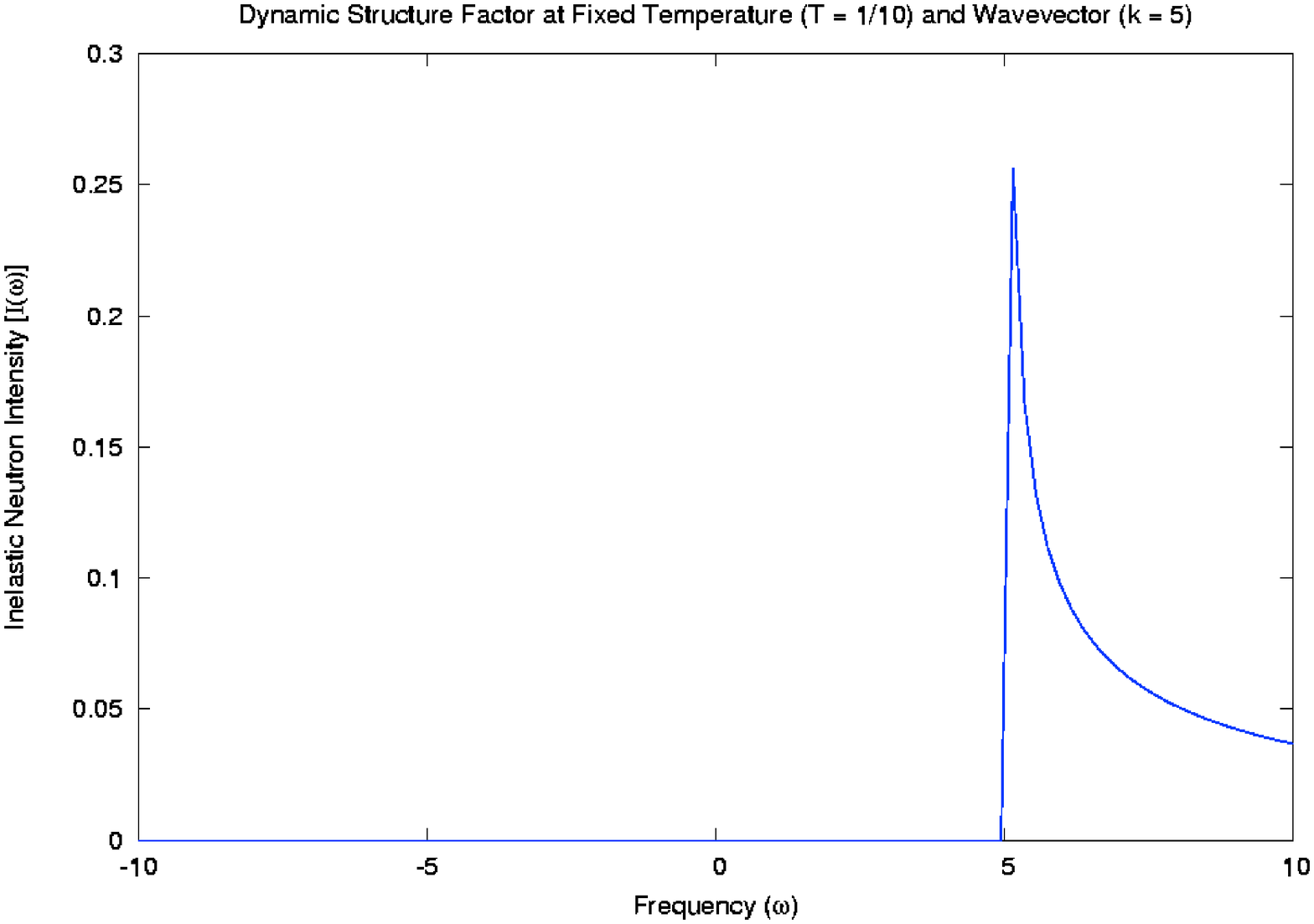}}
	\subfigure{\includegraphics[width=5.5cm,
	    height=4cm]{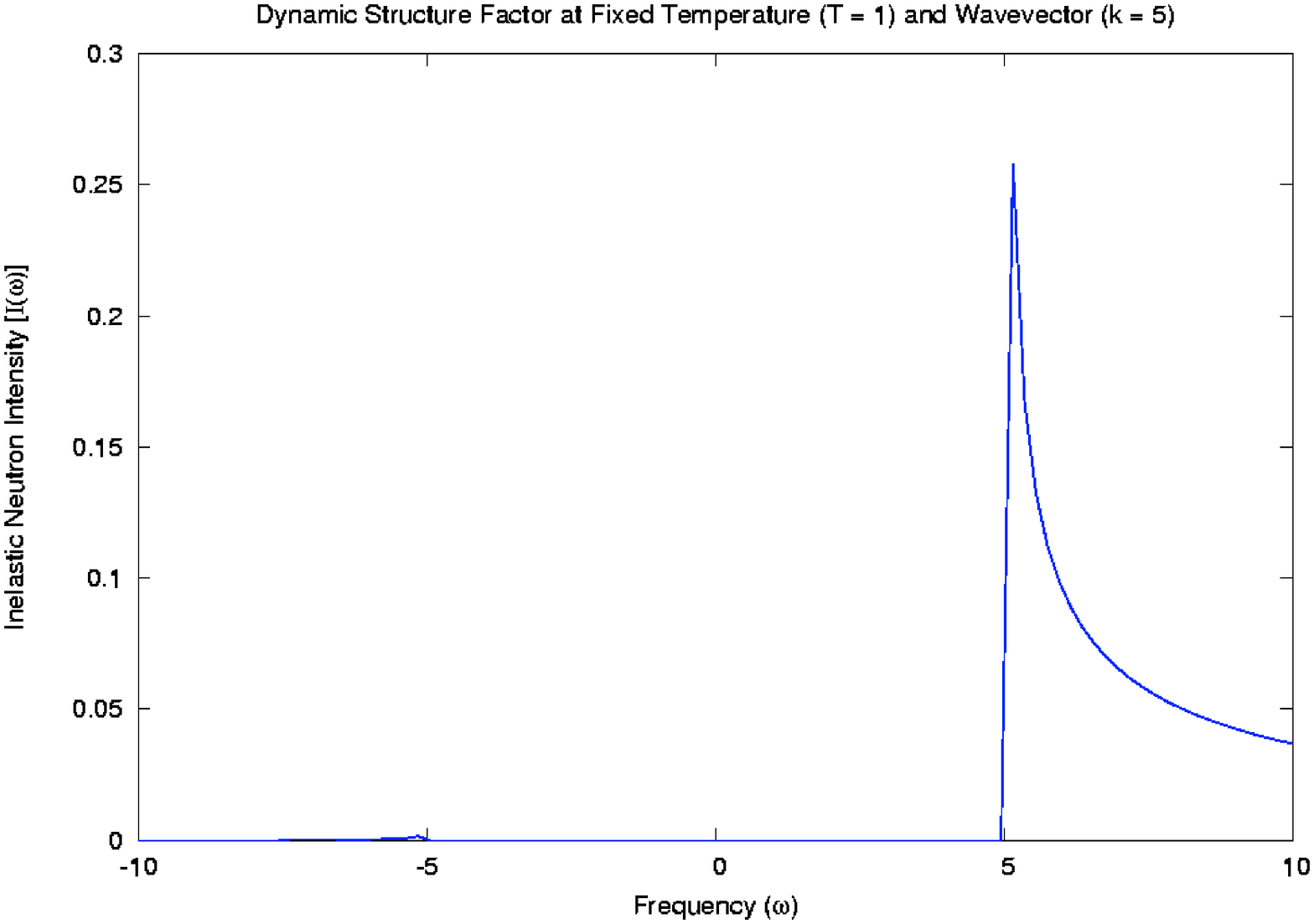}}
	\subfigure{\includegraphics[width=5.5cm,
	    height=4cm]{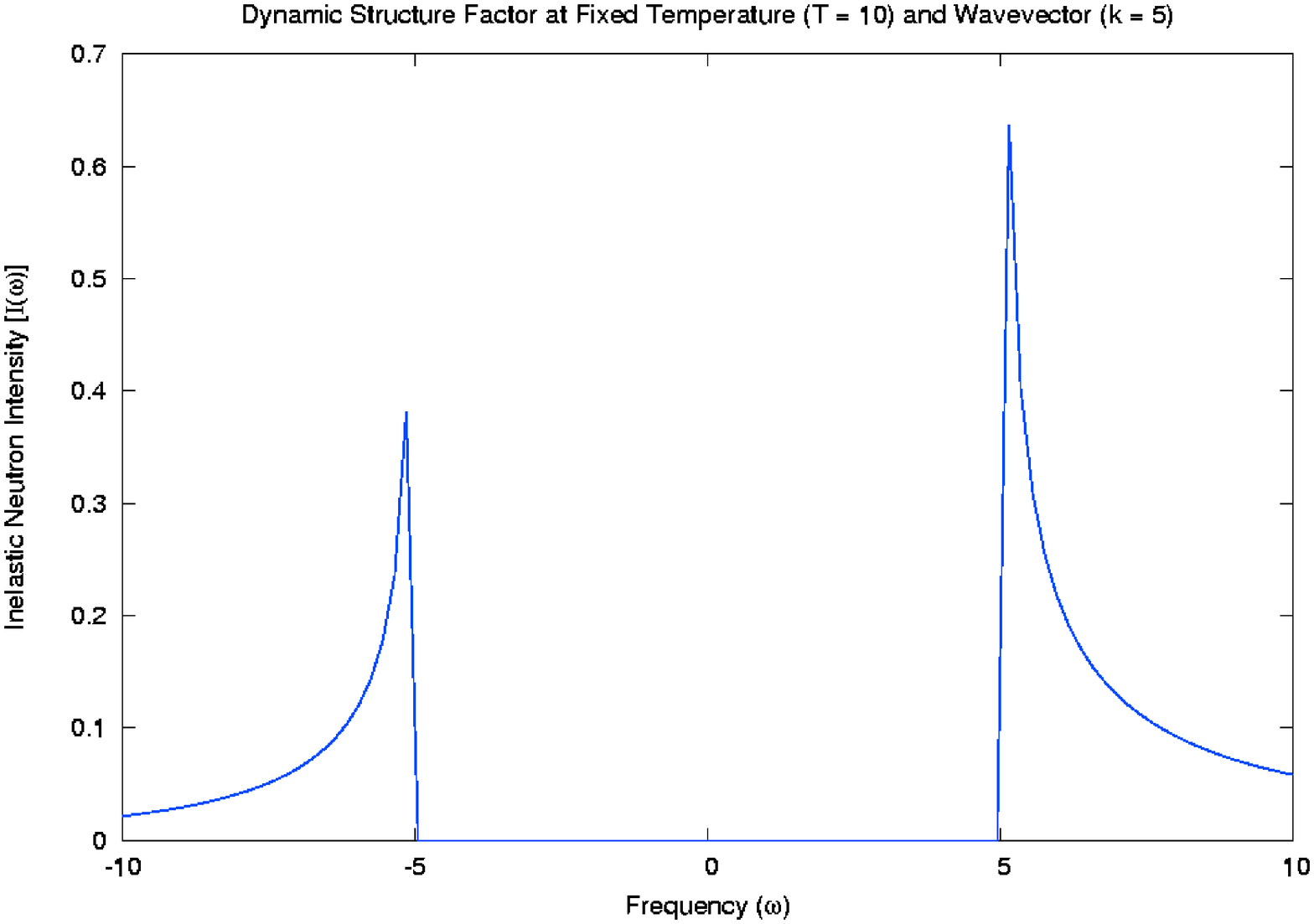}}
	\subfigure{\includegraphics[width=5.5cm,
	    height=4cm]{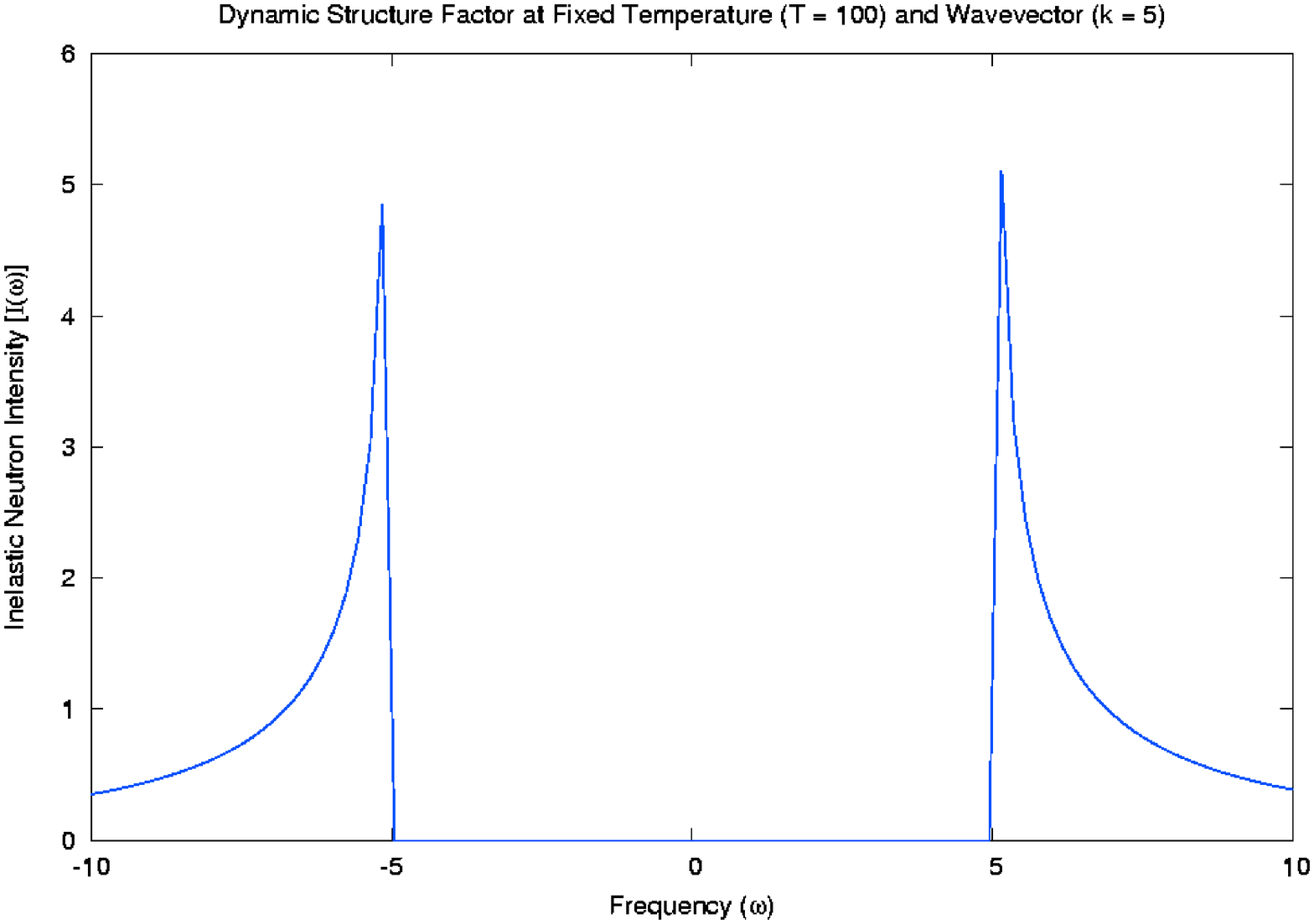}}
	\caption{Inelastic neutron scattering intensity as a function
	  of frequency for fixed momentum and temperatures 1/100, 1/10,
	  1, 10 and 100 respectively.}
	\label{susc}
      \end{center} 
    \end{figure} 

The dynamic structure factor is directly proportional to the density
of states, $S(\omega, \vec k) \propto D(\omega, \vec k, T)$. Hence the
inelastic neutron scattering intensity is given by 
    \beq 
      \nonumber
      \mathcal I (\omega, \vec k) \propto \frac{1}{\pi} \theta \left(
      |\omega| - |\vec k| \right) \frac{1}{\sqrt{\omega^2 - \vec k^2}}
      \, \frac{1}{\left(1 - e^{-\omega / T}\right)} \;.
    \eneq
    \begin{figure}
      \begin{center}
        \subfigure{\includegraphics[width=8.5cm,
        height=6cm]{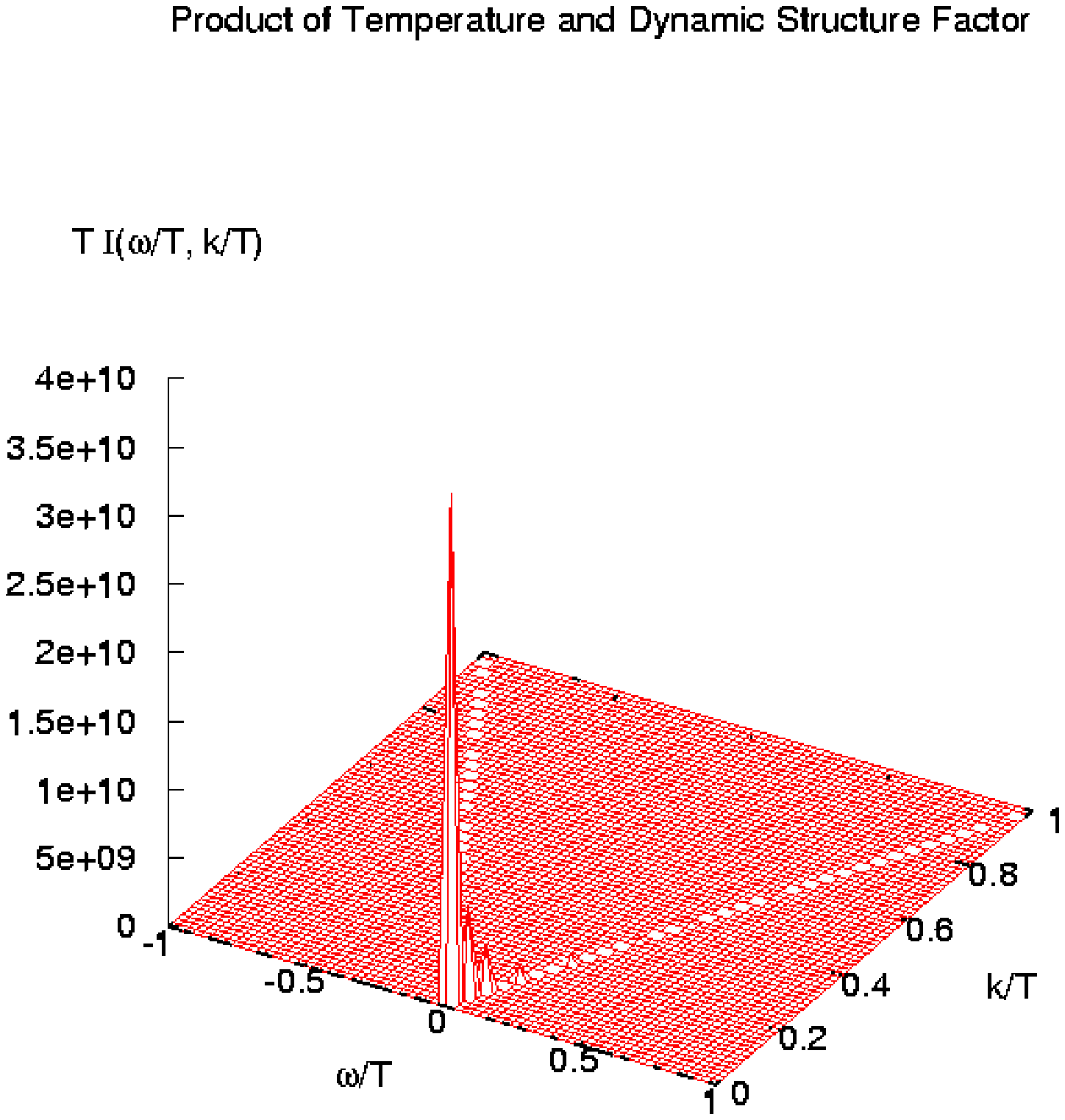}}
	\subfigure{\includegraphics[width=8.5cm,
	height=6cm]{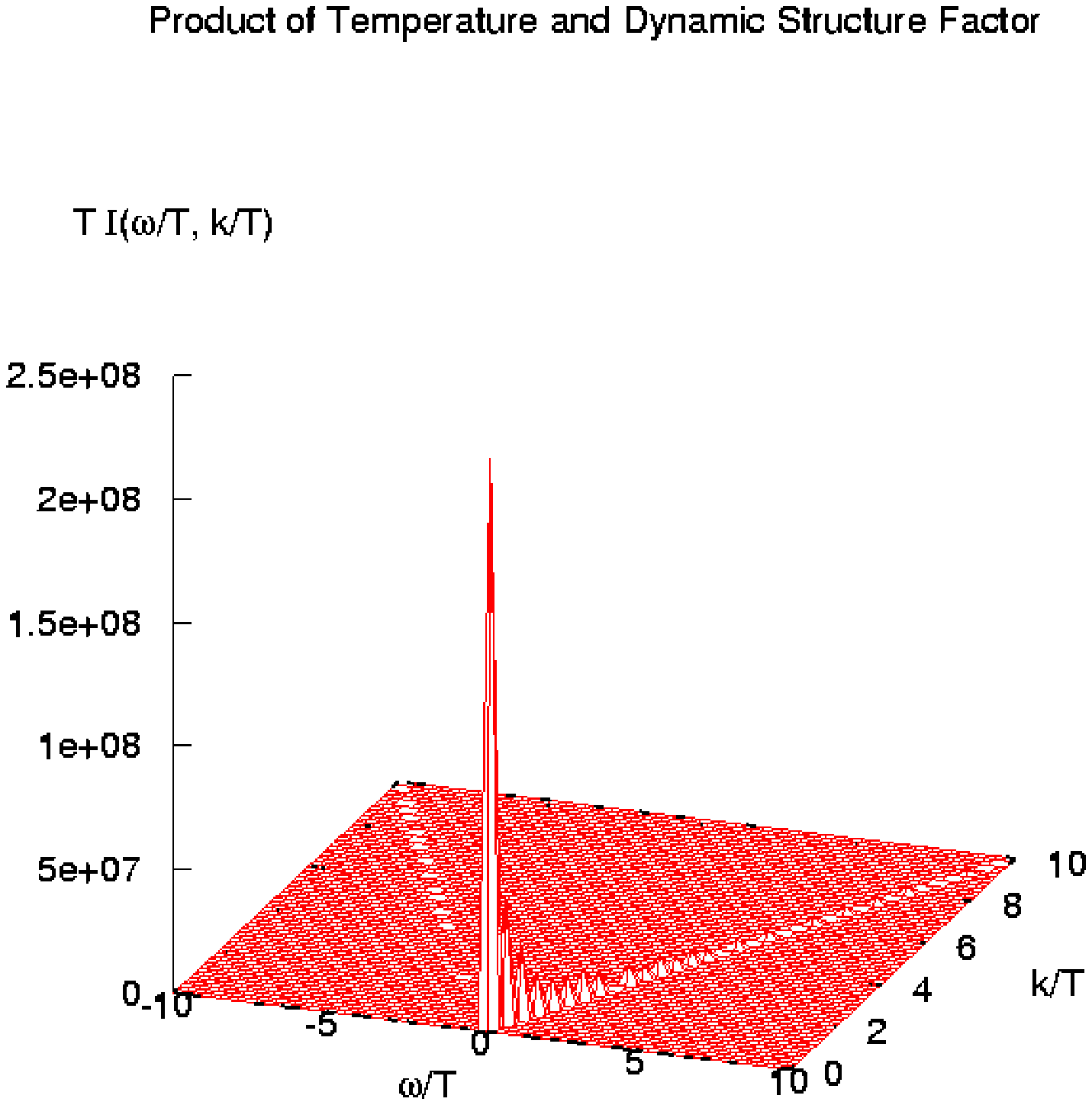}}
	\caption{Scaling of the product of the inelastic neutron
	scattering intensity and temperature with the ratios of
	frequency and momentum to temperature ($\omega/T$ and $k/T$).}
	\label{suscscal} 
      \end{center} 
    \end{figure}
For neutron energy losses which are a lot larger than the temperatures
of the system, the dynamic structure factor is independent of $T$,
with small corrections which are suppressed by exponentials
$e^{-\omega / T}$. For neutron energy losses which are small compared
to the temperatures, the dynamic structure factor is
    \beq 
      \mathcal I (\omega, \vec k) \propto \frac{1}{\pi} \theta \left(
      |\omega| - |\vec k| \right) \frac{1}{\sqrt{\omega^2 - \vec k^2}}
      \, \frac{T}{\omega} \;.
    \eneq 
We plot the neutron intensities as a function of $\omega$ at fixed $T$
and $|\vec k|$ in figure \ref{susc}. The plots can be summarized
because the product of temperature and inelastic neutron scattering
intensity, or structure factor, is a universal scaling function of
frequency over temperature and wavevector over temperature (i.e., it
satisfies a law of corresponding states), which we plot in figure
\ref{suscscal}.

\begin{figure}
      \begin{center}
        \includegraphics[width=5.5cm,height=4cm]{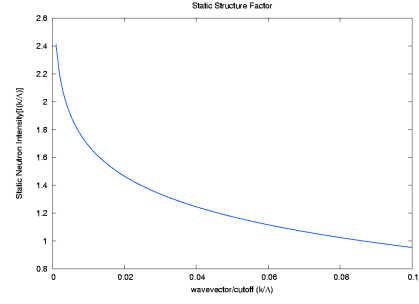}
	\caption{Equal time neutron scattering intensity as a function
	  of the ratio of cutoff to momentum.}
	\label{stat}
      \end{center}
    \end{figure}

Another important measurable quantity is the result of static neutron
scattering experiments. The neutron intensity in this case will be
proportional to the static structure factor. That is, it is
proportional to the inelastic neutron response integrated over all
frequencies
    \begin{align}
      \nonumber
      &\mathcal I (\vec k) \propto \frac{1}{\pi}
      \int_{-\Lambda}^\Lambda d\omega \, \theta \left( |\omega| -
      |\vec k| \right) \frac{1}{\sqrt{\omega^2 - \vec k^2}} \,
      \frac{1}{\left(1 - e^{-\omega / T}\right)} \\
      \nonumber
      &= \frac{1}{\pi} \int_{|\vec k|}^\Lambda d\omega \,
      \frac{1}{\sqrt{\omega^2 - \vec k^2}} \, \left[ \frac{1}{\left(1
      - e^{-\omega / T} \right)} + \frac{1}{\left(1 - e^{\omega / T}
      \right)} \right] \\
      &= \frac{1}{\pi} \int_{|\vec k|}^\Lambda d \omega \,
      \frac{1}{\sqrt{\omega^2 - \vec k^2}} \;.
    \end{align}
The frequency integrals have been cutoff to regularize ultraviolet
divergences. The static neutron intensity is temperature independent
and given by
    \beq 
      \mathcal I (\vec k) \propto \frac{1}{\pi} \ln \left[
      \frac{\Lambda}{|\vec k|} + \sqrt{\frac{\Lambda^2}{|\vec k|^2} -
      1} \right] \simeq \frac{1}{\pi} \ln \left( \frac{2
      \Lambda}{|\vec k|} \right) \;.
    \eneq 
We plot the static neutron intensity as a function of the ratio of
cutoff to wavevector transfer ($\Lambda/|\vec k|$) in figure \ref{stat}.

    \begin{figure}
      \begin{center}
        \subfigure{\includegraphics[width=5.5cm,
        height=4cm]{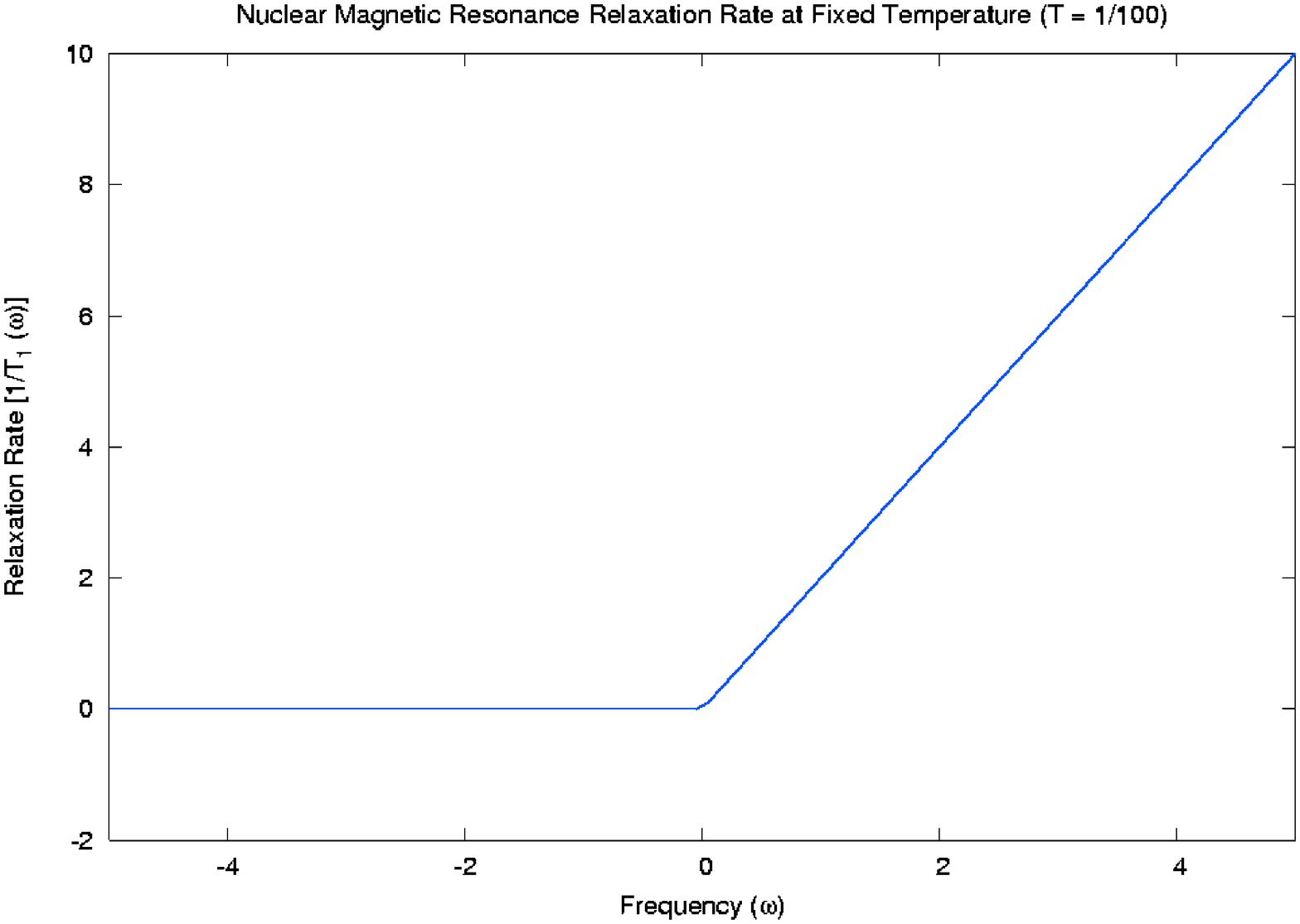}}
        \subfigure{\includegraphics[width=5.5cm,
        height=4cm]{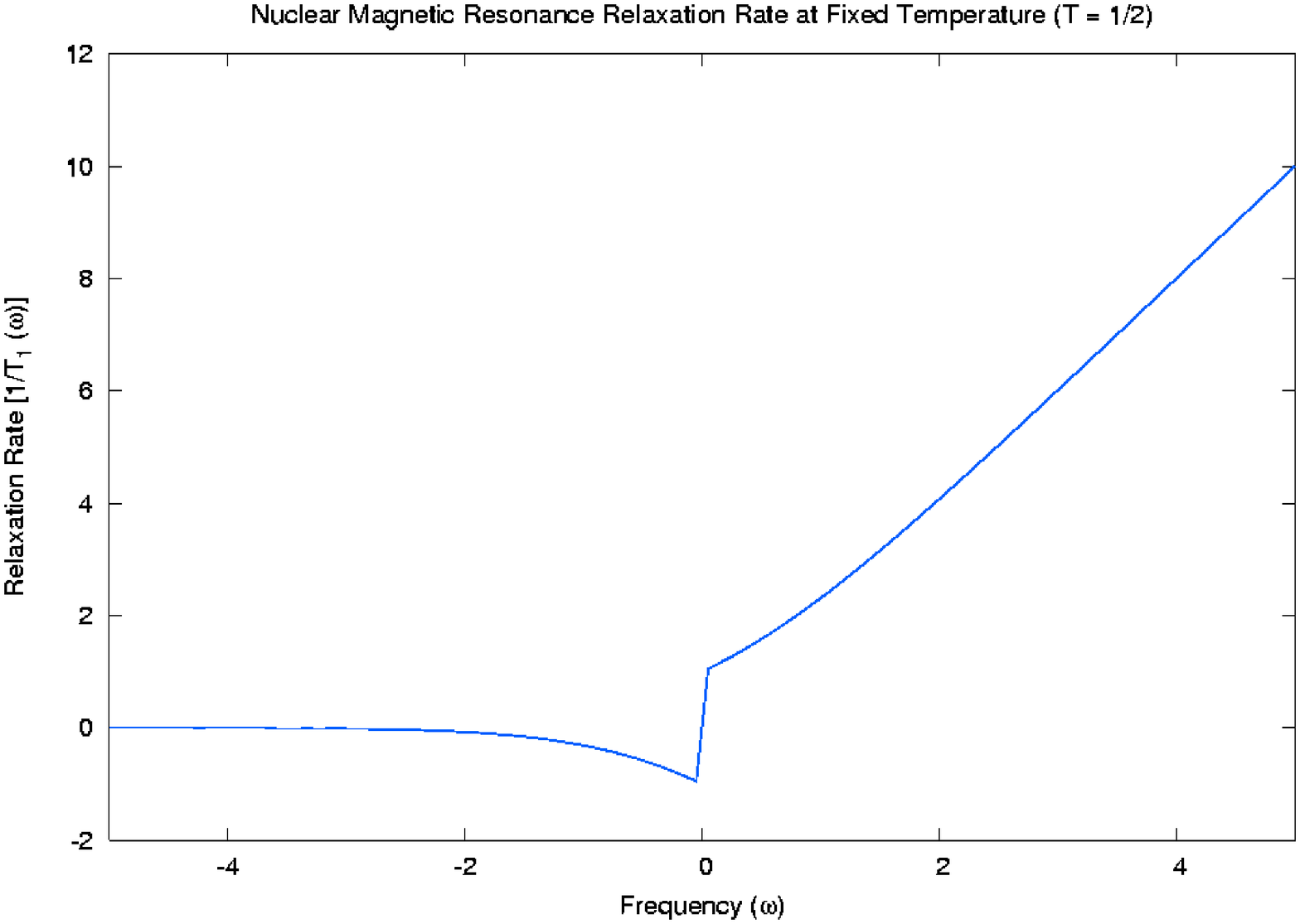}}
        \subfigure{\includegraphics[width=5.5cm,
        height=4cm]{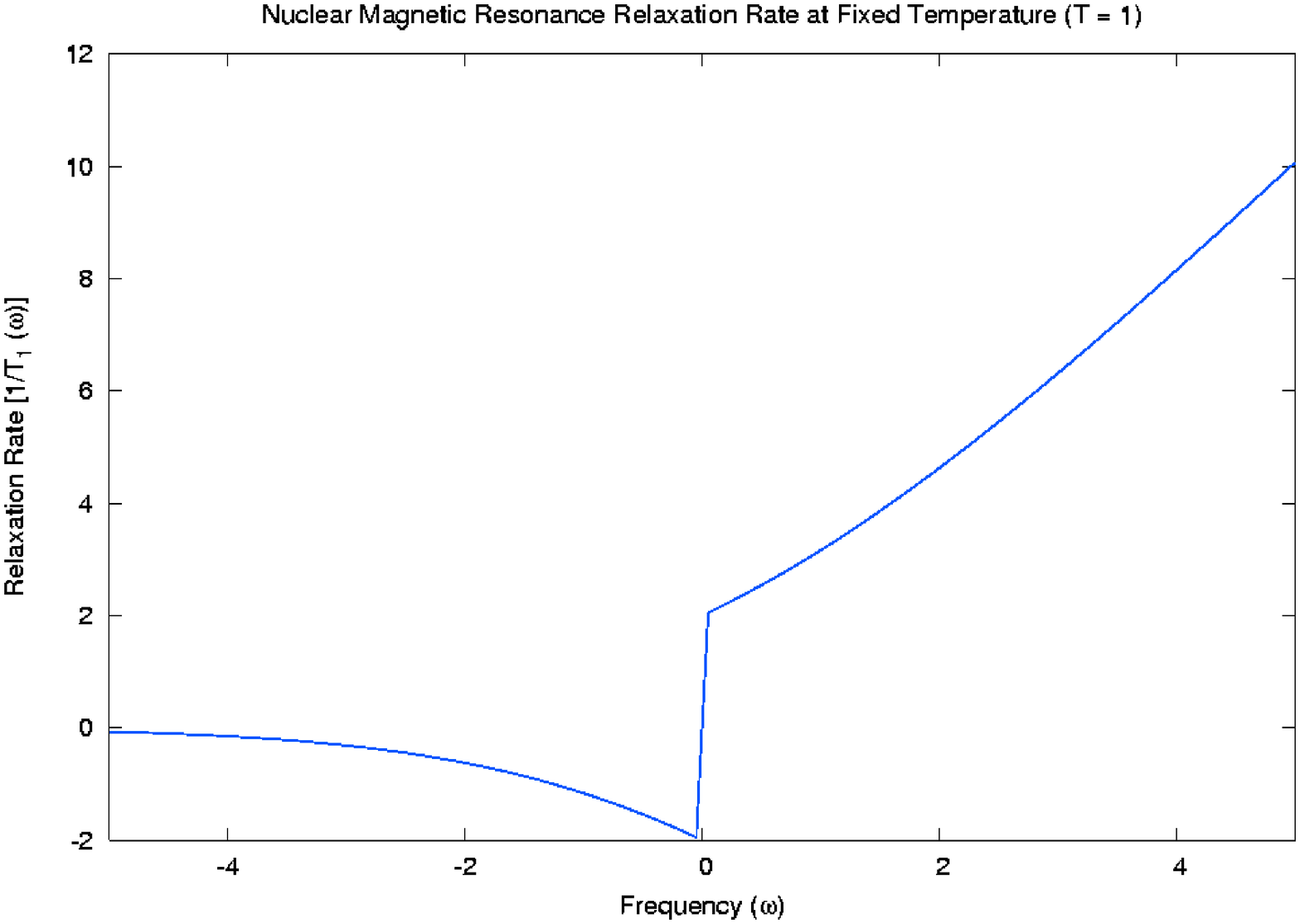}}
        \subfigure{\includegraphics[width=5.5cm,
        height=4cm]{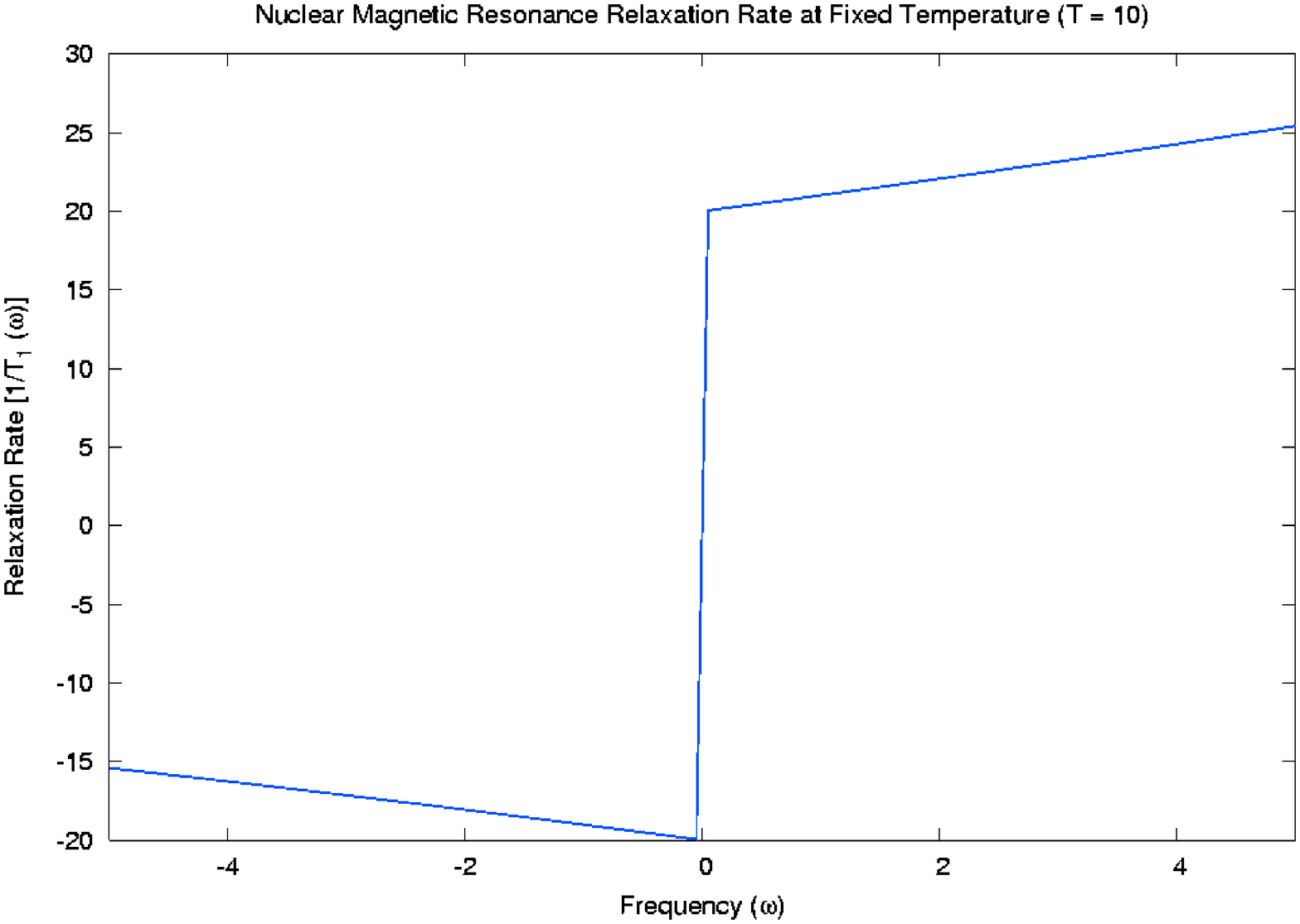}}
        \caption{Nuclear Magnetic Resonance (NMR) Relaxation Rate as a
        function of frequency for temperatures 1/100, 1/2, 1, and 10
        respectively.}
        \label{nmr}
      \end{center}
    \end{figure}

\begin{figure}
      \center
      \includegraphics[width=5.5cm, height=4cm]{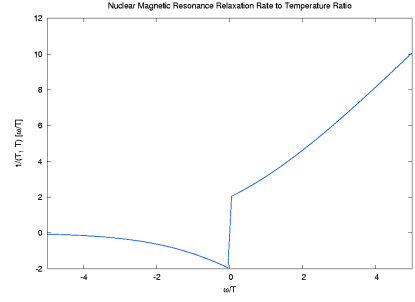}
      \caption{Scaling of ratio of nuclear magnetic resonance (NMR)
      relaxation rate to temperature with $\omega/T$.}
      \label{nmrsc}
    \end{figure}

In some of these antiferromagnetic systems, some of the nuclei making
up the material have nonzero nuclear spin and nonzero hyperfine
coupling to the electronic spins that make up the antiferromagnet. In
these systems one can perform nuclear magnetic resonance (NMR)
experiments. The NMR relaxation rate $1 / T_1$ will be proportional to
the local density of states. Therefore we have
    \begin{align}
      \nonumber
      \frac{1}{T_1} &\propto \frac{1}{\pi} \int d^2 \vec k \, \theta
      \left( |\omega| - |\vec k| \right) \frac{1}{\sqrt{\omega^2 -
      \vec k^2}} \, \frac{1}{\left(1 - e^{-\omega / T}\right)} \\
      \nonumber
      &= \int_0^{|\omega|} d \, |\vec k|^2 \frac{1}{\sqrt{\omega^2 -
      \vec k^2}} \, \frac{1}{\left(1 - e^{-\omega / T}\right)} \\
      &= \frac{2 |\omega|}{1 - e^{-\omega / T}} \;.
    \end{align}
At frequencies small compared with the temperature, the NMR relaxation
rate becomes independent of $\omega$ and linear in temperature. In
fact for $|\omega| \ll T$
    \beq
      \frac{1}{T_1} \propto 2 \, T \, \text{sgn} (\omega) \;.
    \eneq
For frequencies a lot larger than the temperature, the NMR relaxation
rate is proportional to the magnitude of the frequency, and
independent of temperature. In fact, for $|\omega| \gg T$
    \beq
      \frac{1}{T_1} \propto 2 |\omega| \;.
    \eneq
We point out that the NMR relaxation rate, or local susceptibility,
divided by the temperature is a universal scaling function of $\omega
/ T$. We plot the NMR relaxation rate as a function of frequency for
fixed temperature in figure \ref{nmr}.  We summarize the relaxation
rate plots by graphing the scaling plot of the ratio of the relaxation
rates to temperature, which is a universal function of $\omega/T$, in
figure \ref{nmrsc}.

We have just obtained some experimentally measurable response
functions that follow from having an anomalous dimension $\eta=1$ for
the N\'eel magnetization propagator. Since the $CP^1$ mapping of the
N\'eel magnetization into spinons proves quite cumbersome to study the
N\'eel ordered phase, it appears that using our methods we cannot say
much about the critical exponents that follow as the N\'eel phase
perishes. If the critical point where N\'eel order disappears is
indeed a deconfined critical point as the ones studied in the present
work, we can predict relations between the magnetization exponent
$\beta$ and the correlation length exponent $\nu$.

The Josephson correlation length in the N\'eel ordered phase satisfies
the renormalization group (RG) equation\cite{cs1, weinberg, zj, br1,
br2}
    \beq
      \left[ \mu \frac{\partial}{\partial \mu} + \tilde \beta(g)
      \frac{\partial}{\partial g} \right] \xi(\mu,g)=0
    \eneq
with solution
    \beq
      \xi(\mu,g) = \frac{1}{\mu} \exp \left[ - \int_g^{g_c}
      \frac{1}{\tilde \beta(g')} dg'\right]
    \eneq
where
    \beq
      \tilde \beta(g) \equiv \mu \frac{\partial g}{\partial \mu}
    \eneq
is the usual RG beta function and $g_c$ is the critical value of the
coupling constant that separates the N\'eel ordered and paramagnetic
phases: $\tilde \beta(g_c)=0$ and $g_c > 0$. Near the critical point
but on the N\'eel ordered phase, the correlation length scales as
    \beq
      \xi(\mu,g) \sim (g_c - g)^{1/\tilde \beta'(g_c)} \equiv
      \left[\frac{1}{(g_c -g)}\right]^\nu
    \eneq
with $\tilde \beta'(g_c)= d \tilde \beta / dg |_{g=g_c}$. Therefore the
correlation length exponent is
    \beq
      \nu= -\frac{1}{\tilde \beta'(g_c)} \; .
    \eneq

    \begin{figure}[!ht]
      \includegraphics[width=1.3in, height=1.3in]{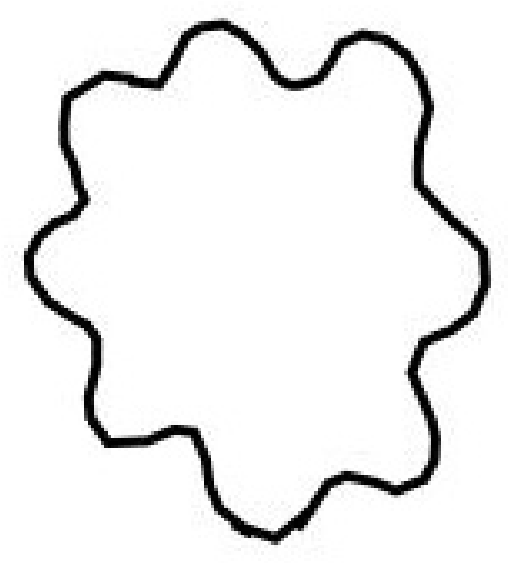}
      \includegraphics[width=1.3in, height=1.3in]{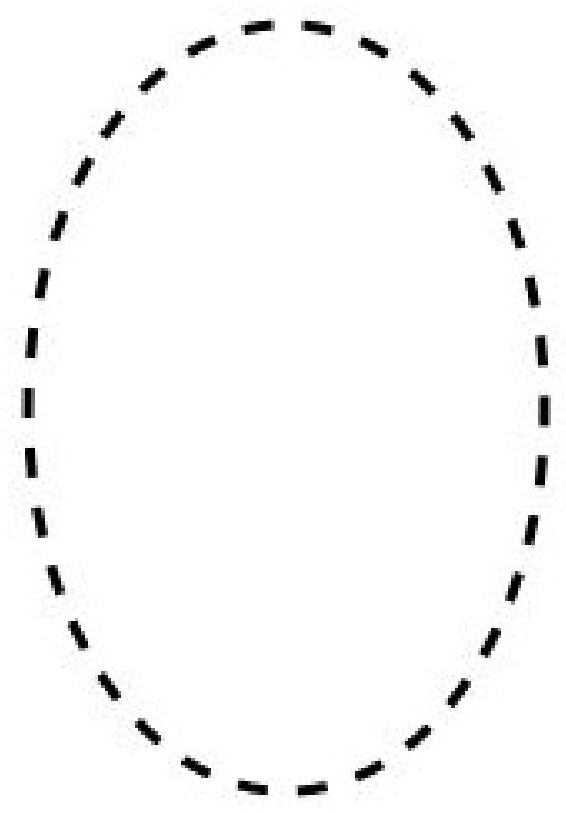}
      \includegraphics[width=1.3in, height=1.3in]{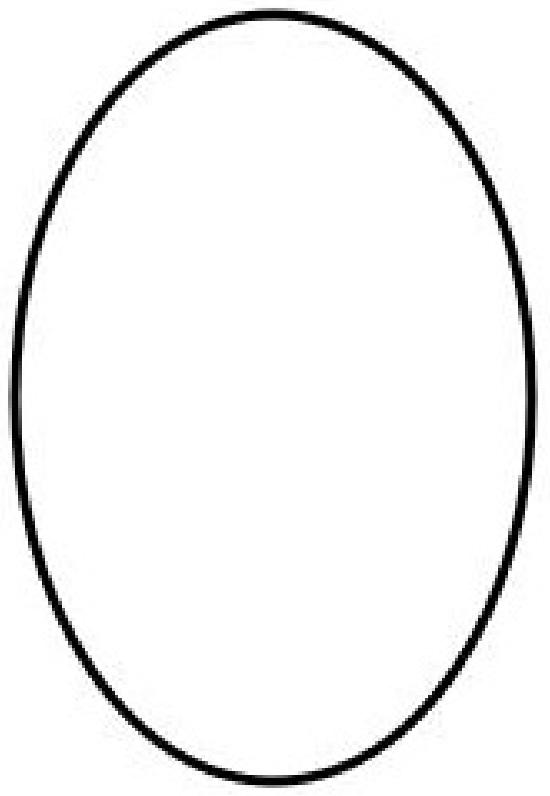}
      \caption{Feynman diagrams representing processes contributing to
      the noninteracting partition function. The diagram on the
      top-left represents the emergent gauge field
      fluctuations. Lagrange multiplier $\delta \lambda$ fluctuations
      are represented by the diagram on the top-right, while spinon
      fluctuations are represented by the diagram on the
      bottom. Wiggly lines represent gauge fields, dashed lines
      represent $\delta \lambda$ fields, and straight lines represent
      spinons.}
      \label{diagZ0}
    \end{figure}

Similarly, the N\'eel magnetization $\sigma$ satisfies the RG equation
\cite{br1,br2,zj}
    \beq
      \left[ \tilde \beta(g)\frac{\partial}{\partial g} + \frac{1}{2}
      \left( 1 + \gamma(g) \right) \right] \sigma(g) = 0
    \eneq
with solution
    \beq
      \sigma(g) = M \exp \left[ -\frac{1}{2} \int_{g_c}^g \frac{\left(
      1 + \gamma(g') \right)}{\tilde \beta(g')} dg' \right] \;.
    \eneq
$M$ is an arbitrary nonuniversal constant and $\gamma(g)$ is the anomalous
dimension obtained from the magnetization renormalization factor $Z$ via
    \beq
      \gamma(g) \equiv \mu \frac{\partial \ln Z}{\partial \mu} \; .
    \eneq
The critical anomalous dimension is then given by $\eta=\gamma(g_c)$,
which is 1 for deconfined spinons. Near the critical point but on the
N\'eel ordered phase the magnetization scales as
    \begin{align}
      \begin{aligned}
	&\sigma(\mu,g) \sim (g_c - g)^{- \left( 1 + \gamma(g_c)
	\right)/[2\tilde \beta'(g_c)]} \\
	&=(g_c - g)^{- \left(1 + \eta \right)/[2\tilde
	\beta'(g_c)]}\equiv (g_c - g)^\beta \;.
      \end{aligned}
    \end{align}
Therefore the magnetization exponent is
    \beq
      \beta = - \frac{\left(1 + \eta \right)}{2\tilde \beta'(g_c)} \;,
    \eneq
which leads to the exponent relation
    \beq
      \beta = \frac{\left(1 + \eta \right) \nu}{2} \; .
    \eneq
For deconfined quantum critical points the correlation length exponent
$\nu$ and the magnetization exponent $\beta$ satisfy the relation
    \beq
      \beta = \nu
    \eneq
which is a unique prediction for deconfined critical spinons.

    \begin{figure}[ht!]
      \includegraphics[width=1in, height=1.5in]{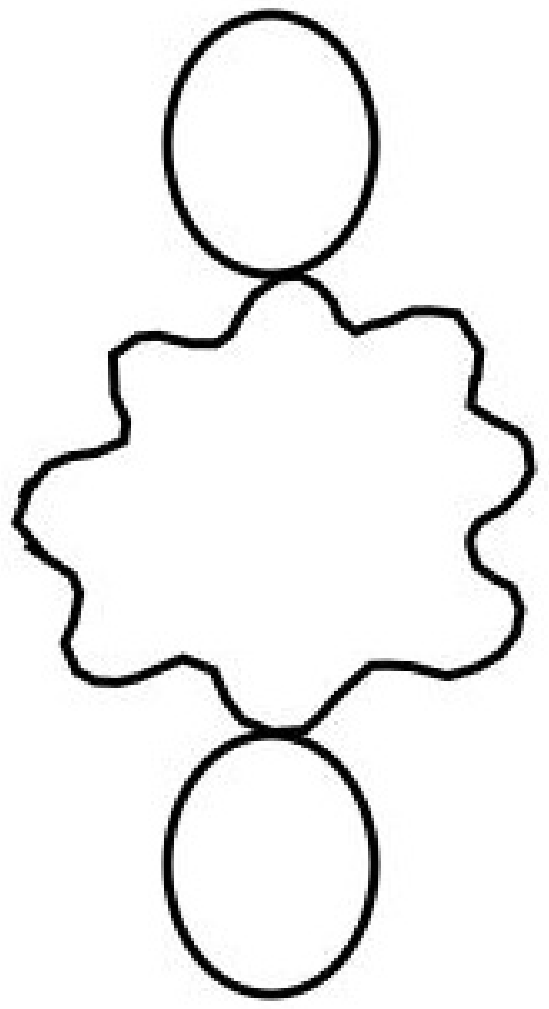}
      \includegraphics[width=1.1in, height=1.5in]{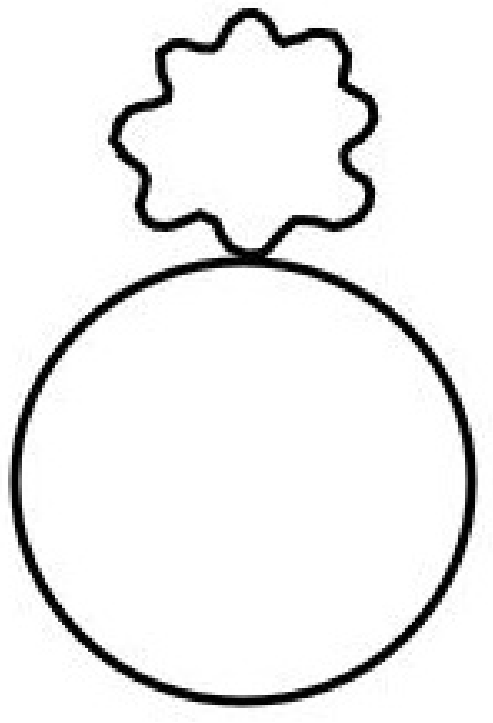}
      \includegraphics[width=1.3in, height=1.3in]{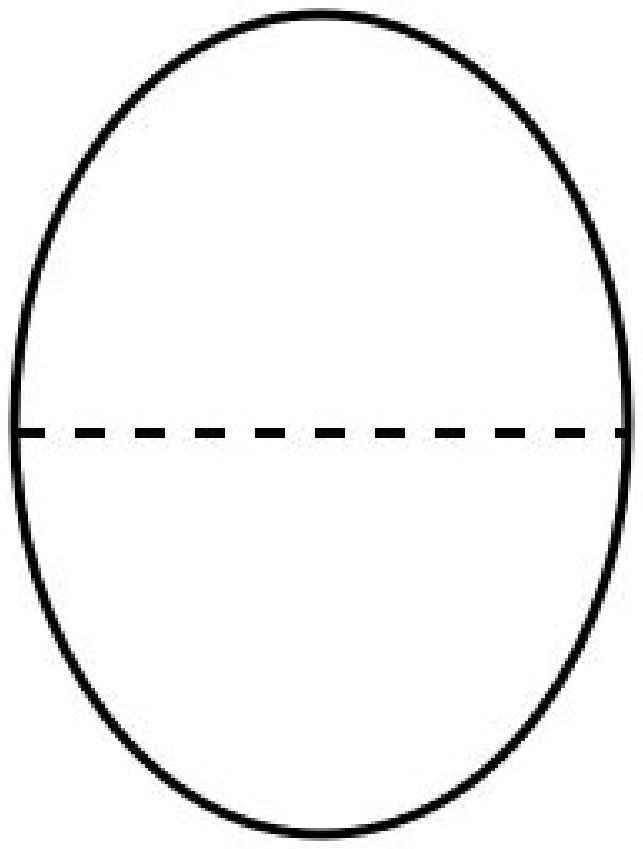}
      \includegraphics[width=1.3in, height=1.3in]{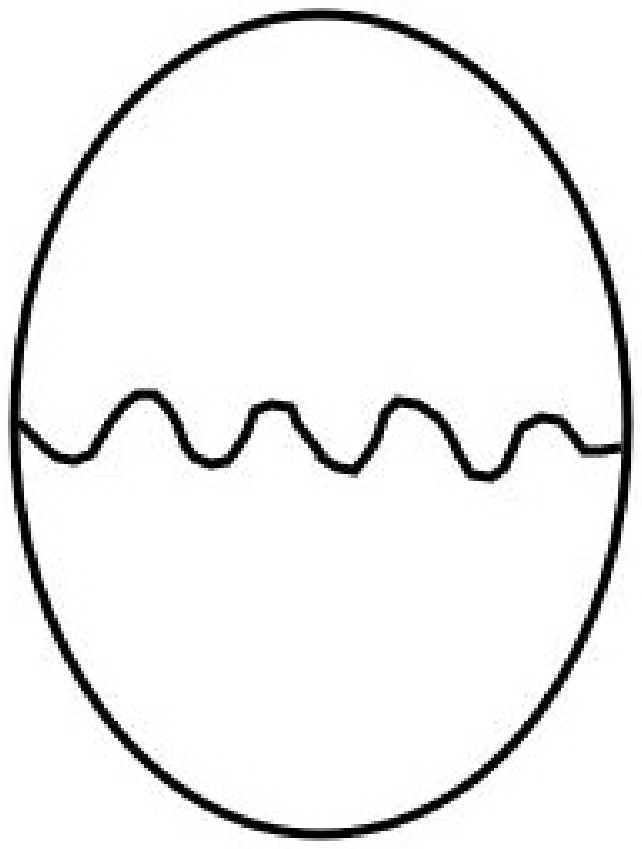}
      \includegraphics[width=1.3in, height=1.3in]{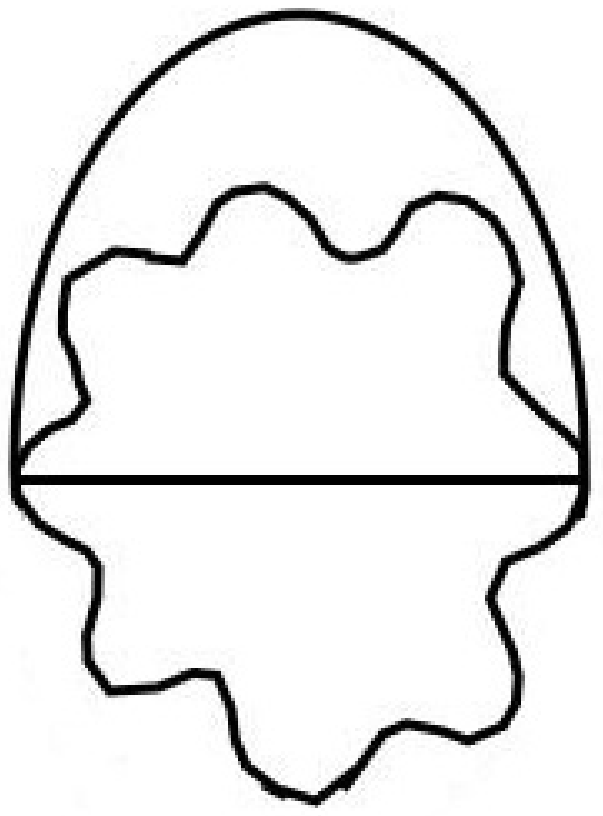}
      \includegraphics[width=0.9in, height=1.9in]{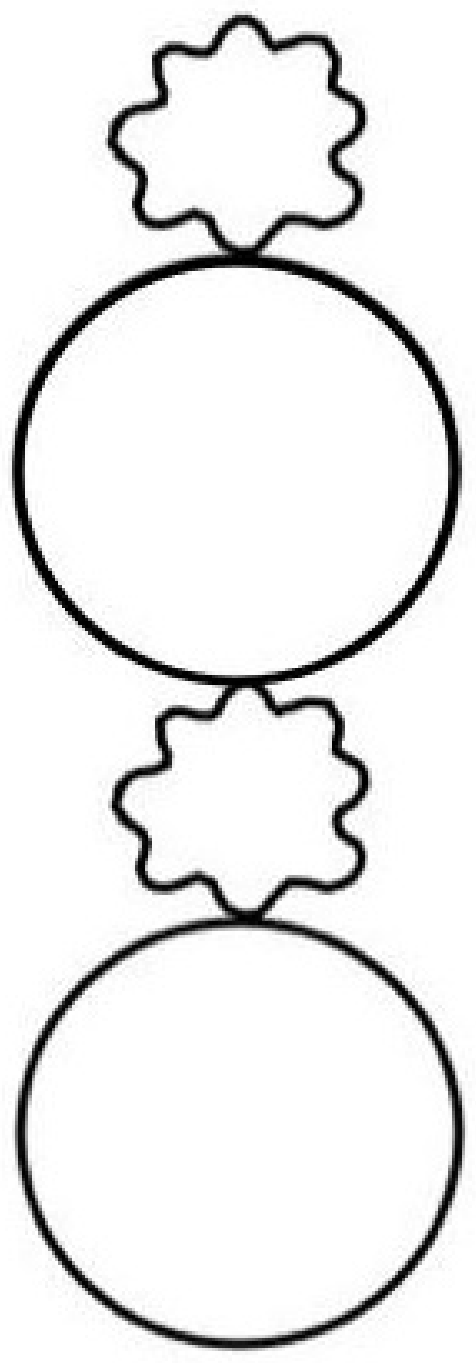}
      \caption{Feynman diagrams representing processes contributing to
      the interacting partition function. These are interactions among
      spinons, Lagrange multiplier fields $\delta \lambda$ and
      emergent gauge fields. Straight lines represent spinons, wiggly
      lines represent emergent gauge fields, and dashed lines
      represent $\delta \lambda$ fields.}
      \label{diagZI}
    \end{figure}

We can also approximate and predict the behavior of the specific heat
at the quantum critical point. Since this is a finite temperature
property, it is usually calculated in Euclidean time, where the
imaginary time direction is finite and goes from 0 to the inverse
temperature $\beta = 1 / T$. The effective action (\ref{seffB}) is
separated into interacting and noninteracting parts. The
noninteracting part includes all quadratic terms. The rest of the
terms are grouped in the interacting part.

Before analysing the action, as with all gauge field theories, we
must fix the gauge for $B_\mu$ in order to remove the gauge
redundancy. This is done by the Faddeev-Poppov procedure, but in this
case in which we will choose the Lorentz gauge, the Faddeev-Poppov
determinant only provides an irrelevant renormalization constant. In
this case, choosing the Lorentz gauge has the only consequence that
the terms $k_\mu B_\mu$ in the action are zero. Hence 
    \beq
      \tilde V_{\mu,\nu}^{-1} (k) \rightarrow \frac{e^2}{\pi^2} \,
      \delta_{\mu\nu} \, k
    \eneq

In order to compute the specific heat, we must compute the free energy
which being the generator of connected Green's functions, it is equal
to $- T \ln Z$, where $\ln Z$ is the log of the free field partition
function plus the sum of connected vaccuum processes. The partition
function receives contributions from the interacting and
noninteracting parts of the effective action. This divides it in a
product of a noninteracting $Z_0$ and interacting $Z_I$ partition
function, such that $Z = Z_0 Z_I$. The relevant Feynman diagrams
contributing to the noninteracting partition function $Z_0$ are shown
in Figure \ref{diagZ0}. These contributions yield a term in the
specific heat which is a positive constant times $T^2$ as expected for
noninteracting relativistic particles.

\begin{figure}[ht!]
      \includegraphics[width=5.5cm, angle=270]{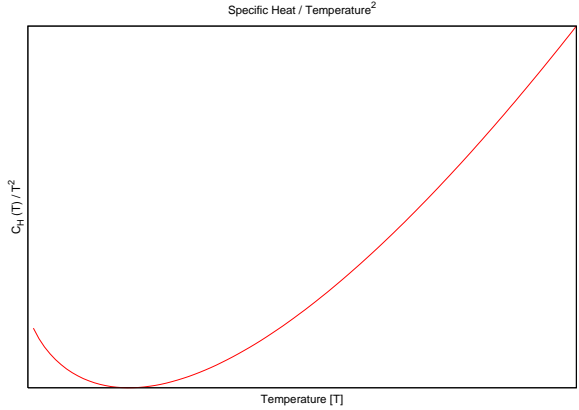}
      \caption{Ratio of the specific heat to $T^2$.}
      \label{CH}
    \end{figure}

The leading contributions to the interacting partition function $Z_I$
are shown in Figure \ref{diagZI}. They give renormalizations of the
$T^2$, noninteracting specific heat.  These contributions also give a
low temperature correction to the specific heat coming from
interactions. Such correction is proportional to a positive constant
times $T^3 \ln T$. The specific heat divided by $T^2$ is plotted in
Figure \ref{CH}.

    \begin{figure}[ht!]
      \includegraphics[width=6.5cm, angle=270]{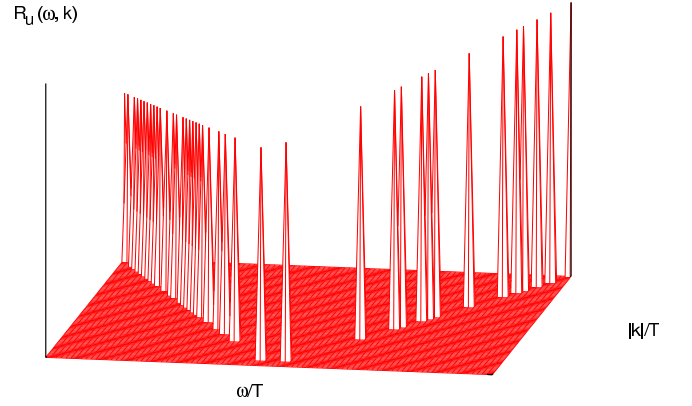}
      \caption{Physical response of the system to an external magnetic
      field as a function of the ratios of frequency and momentum to
      temperature.}
      \label{Rwk}
    \end{figure}

Another interesting experimental signature of deconfined, essentially
free spinons at a quantum critical point is the behavior of the system
in the presence of an external magnetic field. The magnetic field
couples to the N\'eel magnetization vector and rotates it such that
the sigma model partition function in the presence of the external
magnetic field is
    \begin{align}
      \begin{aligned}
	Z &= \int \mathcal D \vec n \, \delta \left( \vec n^2 - 1
	\right) \, e^{- S_E} \\
	S_E &= i S_B + \frac{\rho_s}{2} \int_0^\beta d \tau \int d^2
	\vec x \\
	&\times \left[ \left( \partial_{\vec x} \vec n \right)^2 +
	\frac{1}{c^2} \left( \partial_\tau \vec n - i g_B \vec B
	\times \vec n \right)^2 \right]
      \end{aligned}
    \end{align}

The magnetization is given by 
    \begin{align}
      \begin{aligned}
	M_i &= \frac{\partial \ln Z}{\partial B_i} = \frac{1}{Z}
        \frac{\partial Z}{\partial B_i} \Big |_{\vec B = 0} \\
	&= - \frac{1}{Z} \, \int \mathcal D \vec n \, \delta \left(
	\vec n^2 - 1 \right) \, e^{-S_E} \, \frac{\partial
	S_E}{\partial B_i} \Big |_{\vec B = 0} \\
	&= - \left \langle \frac{\partial S_E}{\partial B_i} \right
	\rangle \Big |_{\vec B = 0}
      \end{aligned}
    \end{align}
and the susceptibility is 
    \begin{align}
      \begin{aligned}
	\chi_{ij} &= \frac{\partial M_i}{\partial B_j} =
	\frac{\partial^2 \ln Z}{\partial B_i \partial B_j} \Big
	|_{\vec B = 0} \\
	&= - \left \langle \frac{\partial S_E}{\partial B_j} \right
	\rangle \left \langle \frac{\partial S_E}{\partial B_i} \right
	\rangle \Big |_{\vec B = 0} \\
	&+ \left \langle \frac{\partial S_E}{\partial B_i}
	\frac{\partial S_E}{\partial B_j} \right \rangle \Big |_{\vec
	B = 0} - \left \langle \frac{\partial^2 S_E}{\partial B_i
	\partial B_j} \right \rangle \Big |_{\vec B = 0}
      \end{aligned}
    \end{align}
The expectation value of the magnetization in the ground state is zero
as it consist of
    \begin{align}
      \begin{aligned}
	&\frac{\partial S_E}{\partial B_i (\vec x, \tau)} =
	\frac{\rho_s}{2 c^2} \, \int_0^\beta d \tau' \int d^2 \vec x'
	\\
	&\times \frac{\partial}{\partial B_i (\vec x, \tau)} \left[ i
	g_B \partial_{\tau'} \vec n (\vec x', \tau') \cdot \left( \vec
	B (\vec x', \tau') \times \vec n (\vec x', \tau') \right)
	\right. \\
	&+ i g_B \left( \vec B (\vec x', \tau') \times \vec n (\vec
	x', \tau') \right) \cdot \partial_{\tau'} \vec n (\vec x',
	\tau') \\
	&+ \left. g_B^2 \left( \vec B (\vec x', \tau') \times \vec n
	(\vec x', \tau') \right) \cdot \left( \vec B (\vec x', \tau')
	\times \vec n (\vec x', \tau') \right) \right] \\
	&= \frac{\rho_s}{2 c^2} \, \left[ i g_B \epsilon_{ijk} \,
	\partial_\tau n_k (\vec x, \tau) n_j (\vec x, \tau) \right. \\
	&+ i g_B \epsilon_{ijk} n_j (\vec x, \tau) \partial_\tau n_k
	(\vec x, \tau) \\
	&+ g_B^2 \epsilon_{ijk} \epsilon_{klm} B_l (\vec x, \tau) n_j
	(\vec x, \tau) n_m (\vec x, \tau) \\
	&+ \left. g_B^2 \epsilon_{jkm} \epsilon_{ilm} B_j (\vec x,
	\tau) n_k (\vec x, \tau) n_l (\vec x, \tau) \right] \\
	&= \frac{\rho_s}{c^2} \left\{ i g_B \, \vec n (\vec x, \tau)
	\times \partial_\tau \vec n (\vec x, \tau) \right. \\
	&+ \left. g_B^2 \, \vec n (\vec x, \tau) \times \left[ \vec B
	(\vec x, \tau) \times \vec n (\vec x, \tau) \right] \right\}
      \end{aligned}
    \end{align}
whose expectation value at $\vec B = 0$ is zero for it consists of
cross products of $\vec n$'s, which make them point in different
directions. The correlation function of two $\vec n$'s pointing along
different directions is zero.

A lengthy but straightforward calculation yields the susceptibility,
which is a nonuniversal constant times $\omega / k$ 
    \beq
      \chi_{ij}^u = \chi^u \delta_{ij} \sim \frac{\omega}{k^2}
      \delta{ij}
    \eneq 
The superscript $u$ differentiates this unstaggered susceptibility
from the N\'eel susceptibility.

The physical response to the magnetic field is proportional to the
imaginary part of the retarded susceptiblity. Use of the fluctuation
dissipation theorem leads to
    \begin{align}
     \begin{aligned}
       R^u (\omega, \vec k) &\sim \frac{1}{1 - e^{- \omega / T}}
       \text{Im } \chi^u (\omega, \vec k) \\
       &\sim \frac{1}{1 - e^{- \omega / T}} \omega \, \delta (\omega -
       |\vec k|) \\
       &- \frac{1}{1 - e^{- \omega / T}} \omega \, \delta (\omega +
       |\vec k|) \\
       &= \frac{1}{1 - e^{- \omega / T}} \frac{\omega}{T} \, \delta
       \left( \frac{\omega}{T} - \frac{|\vec k|}{T} \right) \\
       &- \frac{1}{1 - e^{- \omega / T}} \frac{\omega}{T} \, \delta
       \left( \frac{\omega}{T} + \frac{|\vec k|}{T} \right) \;.
      \end{aligned}
    \end{align}
It exhibits a temperature broadened spinon pole and is thus directly
sensitive to spinon creation. We see that the response is a universal
scale invariant function of $\omega / T$ and $|\vec k| / T$. When the
energy and momentum are just right, the system takes energy from the
magnetic field by shooting off spinons. This behavior is illustrated
in Figure \ref{Rwk}. At zero temperature the response is
    \beq
      R^u (\omega, \vec k) \sim \omega \delta (\omega - |\vec k|) \;.
    \eneq

\begin{figure}[ht!]
      \includegraphics[width=5.5cm, angle=270]{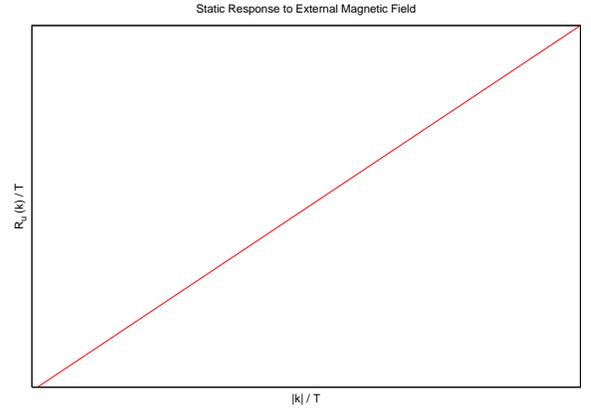}
      \caption{Physical response of the system to an external, static
      magnetic field as a function of the ratio of momentum to
      temperature.}
      \label{Rk}
    \end{figure}

On the other hand, it might be hard to apply a magnetic field with
precisely the right relation between mometum and energy. Thus the
static response, obtained by integrating over all frequencies, might
be of more relevance. It is given by
    \begin{align}
      \begin{aligned}
	&R^u (\vec k) \sim \frac{1}{1 - e^{-|\vec k| /T}}|\vec k| +
	\frac{1}{1 - e^{|\vec k| /T}}|\vec k| \\
	&= T \left ( \frac{1}{1 - e^{-|\vec k| /T}}\frac{|\vec k|}{T}
	+ \frac{1}{1 - e^{|\vec k| /T}}\frac{|\vec k|}{T} \right) \;.
      \end{aligned}
    \end{align} 
and illustrated in Figure \ref{Rk}.

We see that $R^u ( \vec k)/T$ is a universal function of $| \vec k |/
T$. At zero temperature we have
    \beq
      R^u(\vec k) \sim |\vec k| \;.
    \eneq

\begin{figure}[ht!]
      \includegraphics[width=5.5cm, angle=270]{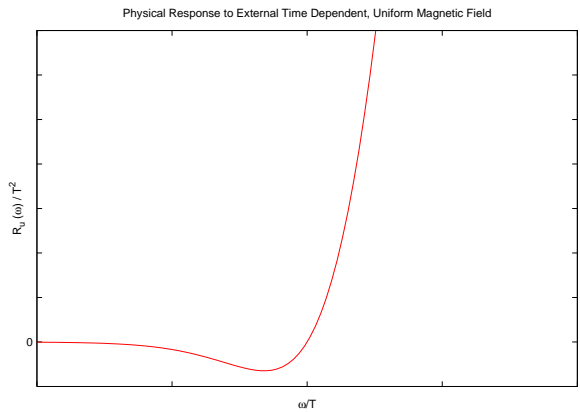}
      \caption{Physical response of the system to an external, time
      dependent, uniform magnetic field as a function of the ratio of
      frequency and temperature.}
      \label{Rw}
    \end{figure}

A similar important quantity is the response to a time dependent but
uniform field, obtained by integrating over all momenta
    \beq
      R^u (\omega) \sim \omega^2 \frac{1}{1 - e^{-\omega /T}} = T^2
      \left( \frac{\omega^2}{T^2} \frac{1}{1 - e^{-\omega /T}}
      \right)\;.
    \eneq
$R^u(\omega)/T^2$ is a universal function of $ \omega/T$ which is
illustrated in Figure \ref{Rw}.
At zero temperature we have
    \beq
      R^u (\omega) \sim \omega^2 \theta(\omega) \; .
    \eneq

We have seen that deconfined quantum critical points are examples of
new types of quantum phase transitions where the standard
Wilson-Ginzburg-Landau criteria that critical properties are
controlled only by order parameter fluctuation fails as has been
recently suggested\cite{sachdev2}. We have uncovered new properties
and physics of such deconfined critical points as well as predicted
some of their experimental features.

\end{document}